\documentclass[twocolumn,amsmath,amssymb]{revtex4} %RevTex 4
\usepackage{graphicx,color}% Include figure files
\usepackage{dcolumn}% Align table columns on decimal point
\usepackage{bm}% bold math
\usepackage{times}
\usepackage{subfigure}
\begin{document}

%\newcommand{\msd}{\mbox{msd}_{\mbox{com}}(t)}

%\definecolor{grey}{rgb}{0.502,0.502,0.502}
%\definecolor{orange}{rgb}{1.,0.5,0.}
%\definecolor{brown}{rgb}{0.55,0.27,0.08}
%\definecolor{dgl}{rgb}{0.,0.3922,0.}

%\preprint{APS/123-QED}

%\title{Manuscript Title:\\with Forced Linebreak}% Force line breaks with \\
%\title{Topology-induced phase separation in semi-dilute solutions of ring polymers}

\title{Numerical study of linear and circular model DNA chains confined in a slit: metric and 
topological properties}

\author{Cristian Micheletti}
\email{michelet@sissa.it}
\affiliation{
SISSA - Scuola Internazionale Superiore di Studi Avanzati and CNR-IOM Democritos,  Via Bonomea 265, 34136 Trieste (Italy)}

\author{Enzo Orlandini}
\email{orlandini@pd.infn.it}
\affiliation{
Dipartimento di Fisica e Astronimia and Sezione INFN,\\ Universit\`a di Padova,
Via Marzolo 8, 35131 Padova (Italy)
}

\date{\today}% It is always \today, today,
 
\begin{abstract}
Advanced Monte Carlo simulations are used to study the effect of nano-slit confinement on metric and topological properties of model DNA chains. We consider both linear and circularised chains 
with contour lengths in the 1.2--4.8 $\mu$m range and slits
  widths spanning continuously the 50--1250nm range. The metric
  scaling predicted by de Gennes' blob model is shown to hold for both
  linear and circularised DNA up to the strongest levels of
  confinement. More notably, the topological properties of the
  circularised DNA molecules have two major differences compared to
  three-dimensional confinement. First, the overall knotting
  probability is non-monotonic for increasing confinement and can be
  largely enhanced or suppressed compared to the bulk case by simply
  varying the slit width. Secondly, the knot population consists of
  knots that are far simpler than for three-dimensional
  confinement. The results suggest that nano-slits could be used in nano-fluidic setups
  to produce DNA rings having simple topologies (including the unknot) or to separate heterogeneous ensembles of DNA rings by knot type.
\end{abstract}

\maketitle

%%%%%%%%%%%%%%%%%%%%%%%%%%%%%%%%%%%%
\section{Introduction}\label{sec:intro}
%%%%%%%%%%%%%%%%%%%%%%%%%%%%%%%%%%%%

Nano-devices are increasingly used to investigate the impact of
spatial restraints on the statics and dynamics of polymers and
biopolymers. DNA is ideally suited for such studies both for its
applicative potential (nano-confinement can be used to sort, sieve and
sequence
DNA~\cite{Han&Craighead:2000:Science,Jo_et_al:2007:PNAS,Zwolak_DiVentra:2008:RevModPhys,Fu_et_al:2006:Phys-Rev-Lett})
and because its elastic properties can be exploited to elucidate
issues of primary interest in polymer
science~\cite{Schoch_et_al:2008:Rev-Mod_Phys,Maier&Radler:1999:Phys-Rev-Lett,Drube_et_al:2010:Nano-Lett,Bonthuis:2008:Phys-Rev-Lett,Lin:2007:Phys-Rev-E,Krishnan_et_al:2007:Nano-Lett,micheletti_physrep}. In
fact, the width of presently-available confining nano-devices (a
channel, a slit, a pit etc.)  can be set to compete with one or more
of the characteristic lengths of a DNA molecule and hence allow for
probing different physical
regimes~\cite{DeGennes:1979,PhysRevE.77.060901,Valle:2005:PRL,Reisner:2005:Phys-Rev-Lett,Ercolini:2007:PRL,Bonthuis:2008:Phys-Rev-Lett,micheletti_physrep,Marenduzzo_et_al_2010_JoP_Cond_Matt,Tang_et_al:2010:Macromol}.

A classic example is offered by the one-dimensional (channel)
confinement of linear DNA molecules whose elongation follows two
different scaling laws, named after de Gennes~\cite{DeGennes:1979} and
Odijk~\cite{Odijik:1983:Macromolecules}, according to whether the
channel width is larger or smaller than $l_p$~\cite{wang2011sim}. A
second notable instance is offered by DNA molecules confined in
three-dimensions, as inside a viral capsid. In the latter case, the
interplay of the genome length and capsid size, can profoundly affect
the entanglement of the packaged
DNA~\cite{Michels&Wiegel:1989:J-Phys-A,Tesi:1998:IMA,Micheletti:2006:J-Chem-Phys:16483240,Tubiana:2011:PRL}. Both
aspects are aptly revealed by the abundance and complexity of knots
that are trapped in DNA filaments that circularise inside viral
capsids~\cite{Arsuaga:2002:Proc-Natl-Acad-Sci-U-S-A:11959991,Arsuaga:2005:Proc-Natl-Acad-Sci-U-S-A:15958528,Matthews:2009:Phys-Rev-Lett:19257792,Marenduzzo:2009:Proc-Natl-Acad-Sci-U-S-A:20018693,Gelbart:2009:Science:19325104,Marenduzzo_et_al_2010_JoP_Cond_Matt,micheletti_physrep,Leforestier:2011:Biophys-J:21539789}.

Compared to the above one- and three-dimensional cases,
two-dimensional polymer confinement is much less characterised both
for the metric and topological properties. To the best of our
knowledge there exist only a limited number of experimental and
numerical studied of the size and shape of DNA molecules in
nano-slits~\cite{Chen_Sullivan:2006:Macromol,Dimitrov_et_al:2008:JCP,Bonthuis:2008:Phys-Rev-Lett,Atapour_et_al:2009:JPA,Tang_et_al:2010:Macromol}. These
seminal studies have addressed the interesting issue of whether
multiple scaling regimes exist in {\em linear} DNA filaments inside
slits. It appears that the same question has not been posed for
circularised DNA molecules where, unlike the linear case, entanglement
is trapped in the form of topological constraints (knots) which affect
the physical behaviour and biological functionality~\cite{Grosberg:2000:PRL,Moore:2004:PNAS,Schvartman_Stasiak:2004:EMBO,Virnau-Mallam-Jackson-knot-review:2011,Virnau:2005:J_Am_Chem_Soc,Sumners_et_al:1995,Liu_et_al:2006:J-Mol-Biol,Liu:2010:J-Mol-Biol:20460130}. The impact of
slit confinement on the topology of circularised polymers has been
previously addressed by early simulations of lattice polymer
models~\cite{Tesi:1994:J-Phys-A} so that its implications for
slit-confined DNA are virtually unexplored.

As a step towards clarifying the above-mentioned aspects we carry out
a systematic numerical study of linear and circularised chains in nano-slits whose salient physical properties, namely the thickness and bending rigidity are set to match those of dsDNA. Specifically we use advanced Monte Carlo techniques to
characterize the shape, orientation and entanglement of linear
and circularised chains of contour lengths in the 1.20--4.8 $\mu$m range
and confined in nano-slits of width from $50$ to $1250$ nm.

The computational scheme is first used to characterize the metric
scaling by varying independently the chain contour length and the slit
width. The metric scaling predicted by de Gennes' blob model is found
to hold, up to the strongest levels of confinement, for both linear
and circular chains.

Next, we show that two-dimensional confinement affects the ring
topology in a qualitatively different way from three-dimensional
confinement.  In fact, for increasing three-dimensional confinement
both the incidence and complexity of knots grows
monotonically~\cite{Micheletti:2006:J-Chem-Phys:16483240,Tubiana:2011:PRL}. In
the two-dimensional case, instead, the knotting probability is
non-monotonic and the knot spectrum mostly consist of the simplest
knot types. The fraction of knotted molecules is maximum, and exceeds
by several times the bulk (unconstrained) value, when the slit width
is a fraction of the average extension of the unconstrained chains. 
A reduction of  the width below this value, causes the knotting probability 
to decrease rapidly.

This fact offers the opportunity to use the set-up of two-dimensional
confinement in nano-slits as an effective method to produce DNA rings
with an {\em a priori} control on their topology and that can be tuned
to yield a proportion of knotted molecules that is much higher or much
lower than the bulk one. In addition, the sensitive dependence of the
conformational entropy of knotted molecules with different topology on
the size of the confining region, suggests that width-modulated slits
could be profitably used in nano-fluidic setups to sort circular DNAs 
by knot type.

\section{Methods}\label{sec:methods}

\subsection{The model}\label{sec:model}

For this study, dsDNA is modelled as a semi-flexible chain of
cylinders. This coarse-grained model captures appropriately the metric
and topological properties of DNA both when it is unconstrained
~\cite{Rybenkov:1993:Proc-Natl-Acad-Sci-U-S-A:8506378,Shimamura&Deguchi:2000:Phys-Lett-A} 
and when it is packed at moderate densities, that is for interstrand separations
larger than $\approx
5$nm~\cite{Micheletti:2008:Biophys-J:18621819}. In such conditions,
which are met in our study, dehydration effects and cholesteric
interactions can be neglected
~\cite{Strey:1998:Curr-Opin-Struct-Biol:9666326,Leforestier:2009:Proc-Natl-Acad-Sci-U-S-A:19470490,Marenduzzo:2009:Proc-Natl-Acad-Sci-U-S-A:20018693,Marenduzzo_et_al_2010_JoP_Cond_Matt,micheletti_physrep}. It
is further assumed that DNA is in a concentrated buffer of monovalent
counterions so that the screened
electrostatic repulsion can also be neglected.
The cylinder diameter is accordingly set equal to the dsDNA hydration
diameter, $d=2.5$nm. 
Notice that no potential energy term is introduced to account for the DNA torsional rigidity. Its interplay with bending rigidity can, {\em a priori} impact the relevant conformational space of densely packed DNA, as argued in ref.~\cite{Grosberg1985} for closed DNA chains. However, recent independent simulations of  three-dimensionally confined DNA with free ends have shown that torsional effects are negligible even at packing densities higher than considered here~\cite{Rollins:2008:Biophys-J:18192353}. We shall therefore work under the hypothesis that the slit-confined DNA chains are torsionally relaxed both in the linear and in the circularised form. For the latter case it is assumed that DNA circularization occurs via non-covalent annealing of the DNA sticky ends so that all torsional stress can still be released.

We consider open and closed chains of contour length equal to $L_c$
ranging from 1.2 $\mu$m to 4.8 $\mu$m and describe them as chains of
cylinders with $N=120 \ldots 480$ vertices, respectively,
as shown in Table \ref{tab:isolchains}. This discretization level
corresponds to a long axis of the cylinder, $b$, approximately equal
to 10nm which is a fraction of the DNA persistence length, $l_p=50$nm.

\begin{table}[htbp]
\begin{center}
\begin{tabular}{|c|c|c|c|c|}
\hline
$ N$ & $ bp $ & $L_c$ [nm] &  Linear $R_g^0$ [nm] & Circular $R_g^0$ [nm] \\
\hline
$ 120 $  &  $ 3534 $  &  $ 1201 $  &  $ 128 \pm 3$ & $ 94 \pm 3 $ \\  
$ 240 $  &  $ 7068 $  &  $ 2403 $  &  $ 190 \pm 5$ &$  136\pm 5 $ \\ 
$ 320 $  &  $ 9426 $  &  $ 3204 $  &  $ 222 \pm 6$ &$  159\pm 6 $ \\ 
$ 360 $  &  $ 10602 $ &  $ 3604 $  &  $ 236 \pm 6$ &$  169\pm 6 $ \\ 
$ 400 $  &  $ 11782 $ &  $ 4006 $  &  $ 251 \pm 6$ &$  179\pm 6 $ \\ 
$ 480 $  &  $ 14136 $ &  $ 4806 $  &  $ 276 \pm 7$ &$  198\pm 7 $ \\ 
\hline
\end{tabular}
\caption{
\label{tab:isolchains}
Parameters of the simulated linear and circular DNA molecules: $N$ is
number of cylinders of the model chains, $bp$ is the total number of base-pairs and $L_c$ is the corresponding contour length. The last two columns provide the bulk values of the root mean square radius of gyration, $R_g^0$ of the linear and circular chains.}
\end{center}
\vspace{-0.6cm}
\end{table}

The potential energy of a chain configuration, $\Gamma$, consists of
two terms, $E_{\Gamma} = E_{excl-vol}(\Gamma)+ E_b(\Gamma)$. The first
term is the excluded volume interaction which enforces the chain
self-avoidance. $E_{excl-vol}$ is set equal to ``infinity'' if two
non-consecutive cylinders overlap, otherwise is set equal to zero.
The second term, $E_b$ is the bending potential energy:
\begin{equation}
E_b = - K_bT \frac{l_p}{b}\sum_{i} \vec t_i\cdot \vec t_{i+1}
\end{equation}
\noindent where $\vec t_i$ is the orientation of the axis of the $i$th
cylinder and the temperature $T$ is set to 300K. For linear chains,
$i$ runs from 1 to $N$. For circular ones, instead, $i$ runs from 1
to $N+1$ with the proviso that $\vec t_{N+1} \equiv \vec t_{1}$. 

\begin{figure*}[htbp]
\includegraphics[width=6.0in]{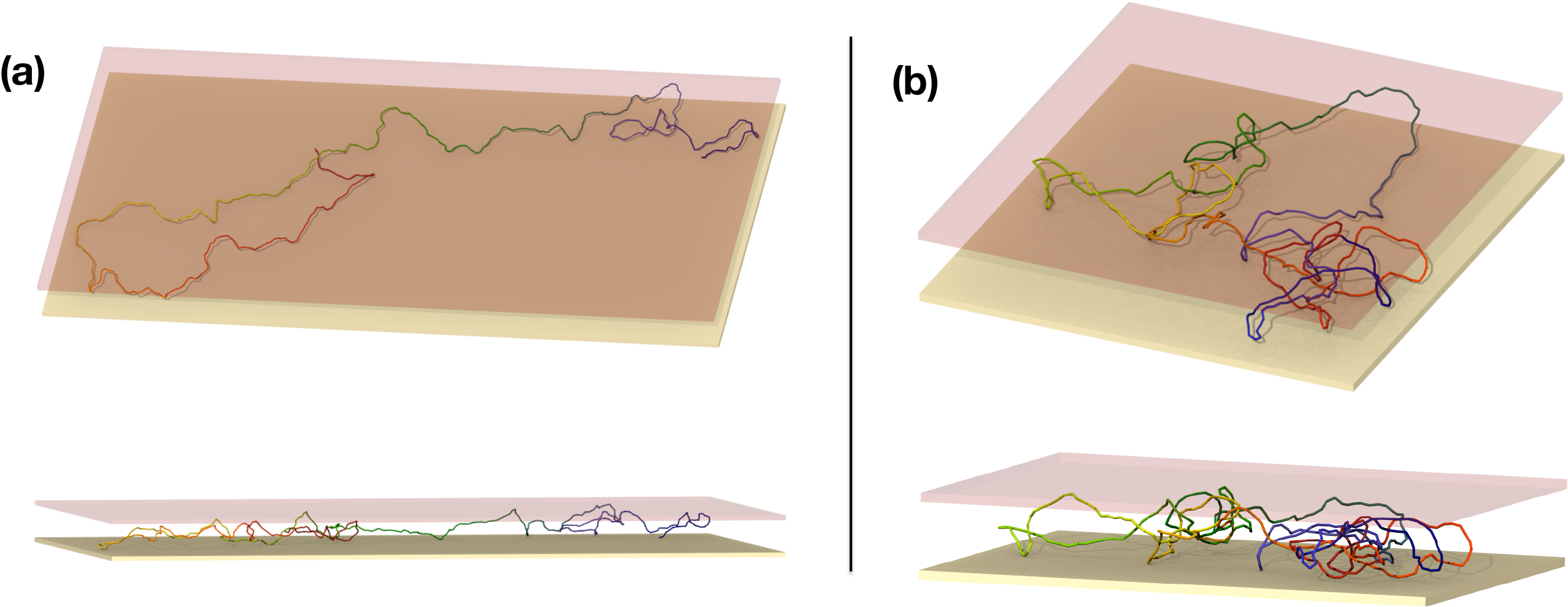}
\caption{Two views of a confined semiflexible linear (panel a) and
  circular (panel b) chains of 480 cylinders, equivalent to a contour
  length $L_c=0.48\mu$m. In both cases the slit width is $D=70$nm.}
\label{fig:slab}
\end{figure*}

DNA confinement in nano-slit is modelled by setting an upper bound,
$D$, to the calliper size of the chains along one direction,
conventionally taken as the $z$ axis of the system Cartesian
coordinate, see Fig.~\ref{fig:slab}. Because the $z$ direction is
perpendicular to the two impenetrable planes delimiting the slit, all
observables measured along this direction will be indicated with the
$\perp$ subscript, while those measured parallel to the slit plane
will be denoted with the $||$ subscript.

\subsection{Advanced sampling techniques}

An advanced Monte Carlo sampling technique is used to compute the
equilibrium metric and topological observables of open and closed
chains subject to slit confinement. The technique is described in
detail in a recent review article~\cite{micheletti_physrep} and
therefore we shall describe it only concisely.

The method consists of using a standard set of crankshaft and pivot
moves to evolve the chains but the canonical weights used for the Metropolis
acceptance criterion are generalised to ensure that configurations are
sampled according to the following (non-normalised) statistical
weight:
\begin{equation}
\exp\left \{-{E_{excl-vol}(\Gamma)+E_b(\Gamma) - \mu \Delta_{\perp}(\Gamma) \over K_b T} \right \}\ .
\end{equation}
\noindent where $\Delta_{\perp}(\Gamma)$, is the configuration
calliper size measured along the confining, $z$ direction.  The
non-negative quantity $\mu$ can be viewed as an anisotropic pressure
promoting the confinement along $z$. In fact, by increasing $\mu$ one
can reduce the average calliper size of the generated configurations.
By using a multiple Markov chain scheme, consisting of several
parallel Monte Carlo simulations each at a different value of
$\mu$~\cite{Tesi_et_al:1996:JSP,Micheletti:2006:J-Chem-Phys:16483240},
it is possible to sample very effectively configurations covering a
wide range of calliper size, $\Delta_{\perp}(\Gamma)$. Finally, a
thermodynamic reweighting technique analogous to the one used in
ref.\cite{Micheletti:2006:J-Chem-Phys:16483240} is used to combine
the observables computed for configurations generated at various
values of $\mu$. This procedure gives an estimate of the equilibrium properties of chains that
can be accommodated inside slits of width, $D$, in the 50nm -
1250nm range.

\subsection{Metric properties}

The average size, shape anisotropy, and orientation of a given ring
are characterized by means of the eigenvalues and eigenvectors of its
gyration tensor, ${\mathbf Q}$. The entries of this $3 \times 3$ symmetric
matrix are given by:
\begin{equation}\label{eq:gyrtensdef}
Q_{\alpha, \beta} =
\frac{1}{N} \sum_{i=1}^{N}
({\bf r}_{i, \alpha} - \bar {\bf r}_{\alpha}) ({\bf r}_{i, \beta} - \bar{\bf r}_{\beta}) ,
\end{equation}
where ${\bf r}_{i}$ is the position of the $i$th vertex of the chain,
$\alpha$ and $\beta$ run over the three Cartesian components and
$\bar{\bf r} = \frac{1}{N} \sum_{i} {\bf r}_i$ is the
average vertex position in the chain. The non-negative eigenvalues of
${\mathbf Q}$, ranked for decreasing magnitude, are indicated as $\Lambda_1$,
$\Lambda_2$ and $\Lambda_3$ and their associated (normalised)
eigenvectors are indicated as $\vec{e}_1$, $\vec{e}_2$ and $\vec{e}_3$.

The sum of the eigenvalues provides the square radius of gyration of
the chain:
\begin{equation}
R_g^2 = \Lambda_1 + \Lambda_2+ \Lambda_3\ ,
\end{equation}
\noindent while the adimensional ratios $\Lambda_1/\Lambda_3$ and
$\Lambda_2/\Lambda_3$ are used to characterize the chain
shape anisotropy.

The chain orientation inside the slit is conveniently described by the angle
$\alpha$ formed by each principal gyration axis and the slit plane:
\begin{equation}
\cos \alpha_i = \sqrt{1 - | \vec{e_i} \cdot \hat{\bf z}|^2}
\label{eqn:alphai}
\end{equation}
\noindent where $\hat{\bf z}$ is the unit vector in the $z$ direction.

To parallel the quantities that are customarily measured in nano-slit
experiments, for each chain configuration we compute its mean square
size projected in and out of the slit plane which are, respectively:

\begin{eqnarray}
R^2_{||} &=& \sum_{i=1,2,3} \Lambda_i \cos^2 \alpha_i,\\
R^2_{\perp} &=& R^2_g - R^2_{||} = \sum_{i=1,2,3} \Lambda_i (1-\cos^2 \alpha_i) \ .
\end{eqnarray}

\subsection{Scaling properties and the deGennes regime}

The de Gennes' blob theory provides a powerful reference framework for
characterizing spatially constrained chains~\cite{DeGennes:1979}.  
The theory, originally formulated to improve on standard mean-field theories for semi-dilute polymer solutions and~\cite{deGennes1975a,deGennes1975b}, has proved useful to understand the behaviour of polymers confined in
regions with transverse size larger than the molecule's persistence
length.

\begin{figure}[htbp]
\includegraphics[width=3.0in]{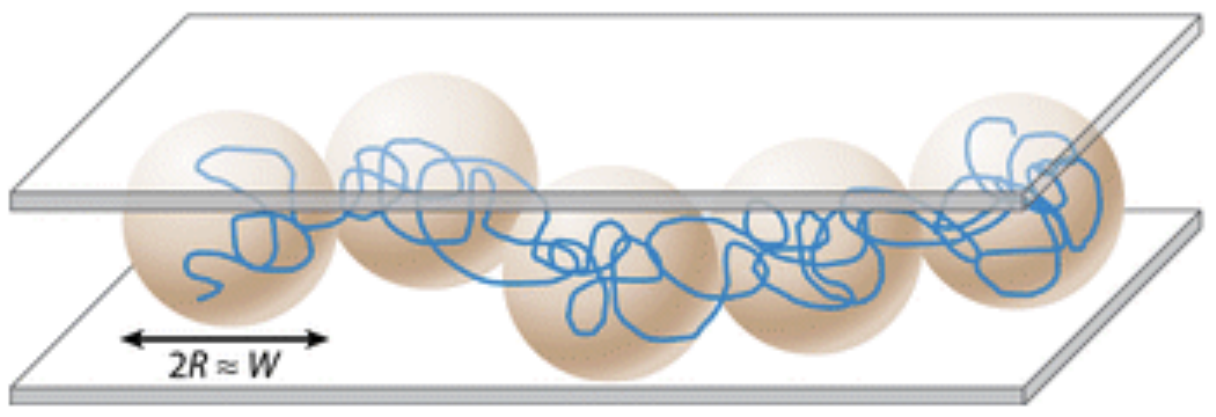}
\caption{Pictorial representation of a confined polymer as a string of
  DeGennes' blobs}
\label{fig:blobs}
\end{figure}

When such condition is met, the theory of de Gennes proposes to describe
the self-avoiding chain, formed by $N$ monomers of size $b$, as a
sequence of ``blobs'', with each blob consisting of $g$ monomers and
having linear size about equal to the slit width, $D$, see
Fig.~\ref{fig:blobs}. Within each blob, the chain does not experience
the confining constraints and hence it behaves as an unconstrained
self-avoiding walk: $D \sim g^{\nu_{\rm 3D}}$, where $\nu_{3D}\sim
0.588$ is the three-dimensional self-avoiding walk scaling
  exponent~\cite{Clisby:2010:Phys-Rev-Lett}. At the same time, the
  string of blobs can be viewed as being a two-dimensional self-avoiding
  walk, with the blobs being the effective monomeric units of the
  walk. Accordingly, the in-plane size of the chain expectedly scales as
  $R_{||} \sim (N/g)^{\nu_{\rm 2D}} D$ where $\nu_{\rm 2D}=3/4$ is the
  self-avoiding exponent in two dimensions.

This relationship can be  recasted as
$R_{||} \sim (N/g)^{\nu_{\rm 2D}} D \sim N^{\nu_{\rm 2D}} D^{-\nu_{\rm 2D}/\nu_{\rm 3D}} D$ or, equivalently,
\begin{equation}
{R_{||} \over R_{||}^0} \sim {\left(D \over R_g^0 \right)}^{ 1 - \nu_{\rm 2D}/\nu_{\rm 3D}} \sim  {\left( D \over R_g^0 \right)}^{-0.276}\ ,
\label{eqn:dgs}
\end{equation}

\noindent where we have exploited the proportionality of $R_g^0$ and
$R_{||}^0$ which are respectively the root mean square values of the
radius of gyration and the in-plane projected size of unconstrained
chains. 
It is worth pointing out that the prefactor entering the scaling relationship of eq.~\ref{eqn:dgs} changes according to whether the blobs are isometric or not~\cite{PhysRevE.77.060901,wang2011sim}. This condition depends, in turn, on whether the slit width $D$ is larger or smaller than $l_p^2/d$. 
For our case, the latter quantity is equal to 1$\mu$m and therefore is always larger than the widths of the slits considered. 
The blobs are therefore expected to be anisometric 
(extended de Gennes regime)~\cite{wang2011sim}.

\subsection{Topological properties}

The characterization of the entanglement of circular configurations is
completed by establishing their knotted state. It should be noted, in
fact, that the crankshaft moves employed for the MC evolution preserve
the ring connectivity but not necessarily its
topology~\cite{Orlandini&Whittington:2007:Rev-Mod_Phys}. In fact, even
if the initial and final state of a crankshaft move are self-avoiding,
the virtual crankshaft movement that bridges them can entail several
bond crossings and hence can change the ring knotted state.

The topology of a ring is established using the same methodology of
ref. \cite{Micheletti:2006:J-Chem-Phys:16483240,Micheletti:2008:Biophys-J:18621819}
which is articulated over the following steps. First, the ring
geometry is simplified by topology preserving
moves~\cite{Taylor:2000:Nature:10972297,Koniaris&Muthukumar:1991a,Micheletti:2006:J-Chem-Phys:16483240,micheletti_physrep}. These
moves consist of ring deformations involving no bond crossing and
promote the collinearity of triplets of ring vertices that are nearby
in sequence. After the chain simplification, a two-dimensional
projection of the resulting configuration is next taken, encoded as a
digital Dowker code, simplified algebraically and finally compared
against a lookup table of prime knots with up to 16 crossings. If a
match is found for all the prime component of a knot, then the ring knotted
topology is unambigously established otherwise it is labelled
as ``unknown''.

Notice that the ``unknown'' states may correspond to complicated knots
with prime components having more than 16 crossings, or may be a
simple knot that could not be simplified enough to positively identify
it.

Further insights on the interplay between the topological properties
of circular DNAs and the degree of confinement may be obtained by 
considering the average length of the knotted region in the ring.
Locating the knotted portions in closed chains is a challenging
problem, that presents the same conceptual difficulties as
establishing the knotted state of a geometrically-entangled open
chain~\cite{Orlandini&Whittington:2007:Rev-Mod_Phys,micheletti_physrep}.
Although the application of topological concepts to open chains is
{\em a priori} ambiguous, it has been recently verified that very
different methods for locating knots in linear chains yield remarkably
consistent results~\cite{Min_entang_closure}. Among them we mention
those based on the statistical closure scheme~\cite{Millett:2005:Macromol} and the minimally-interfering one~\cite{Min_entang_closure} which is used here for its robustness and
numerical efficiency.

\section{Linear DNA in nano-slits}\label{sec:linear}

\subsection{Effect of confinement on chain orientation and shape}

\begin{figure}[h!]
\centering
\subfigure{
\includegraphics[width=3.0in]{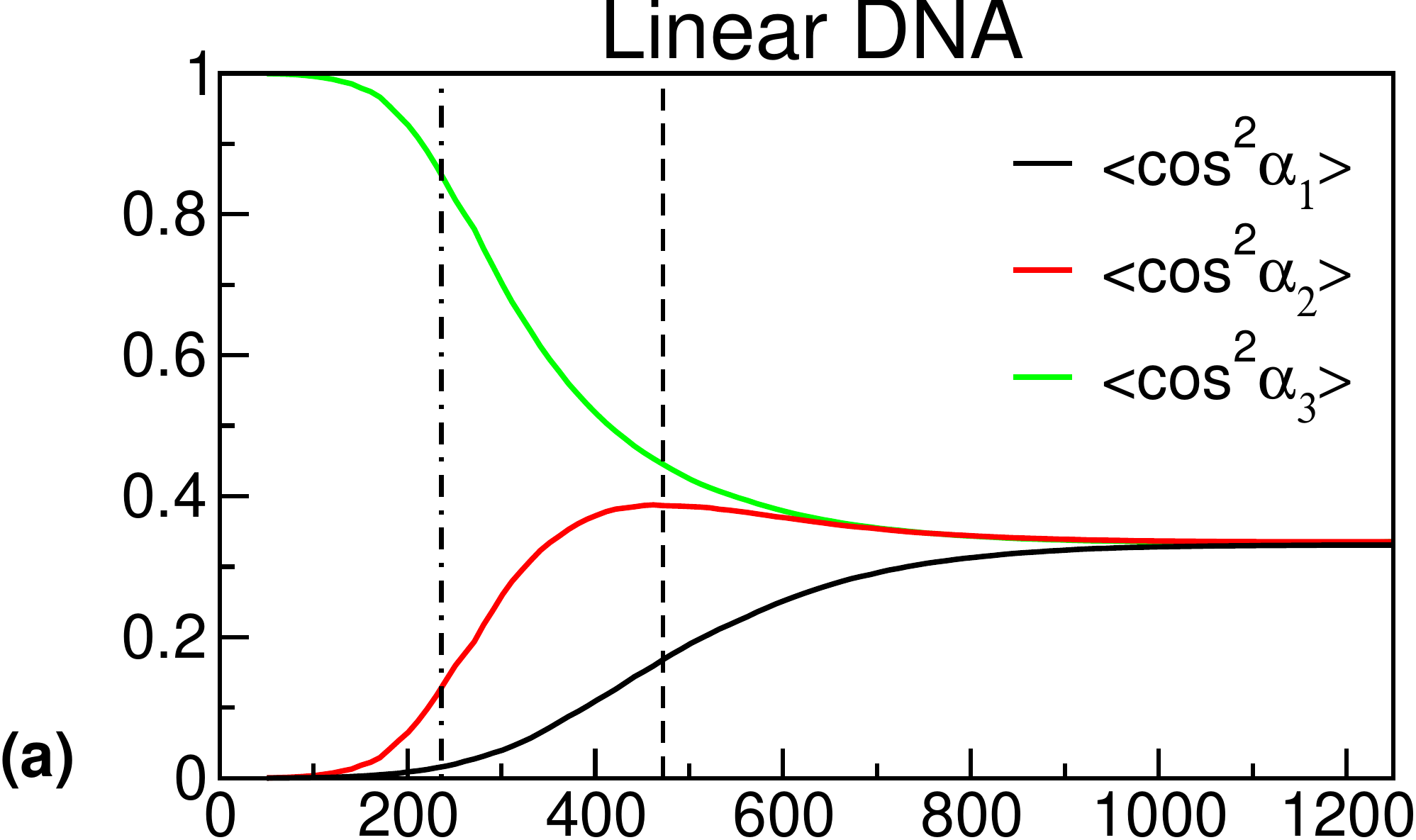}
}\\
\subfigure{
\includegraphics[width=3.0in]{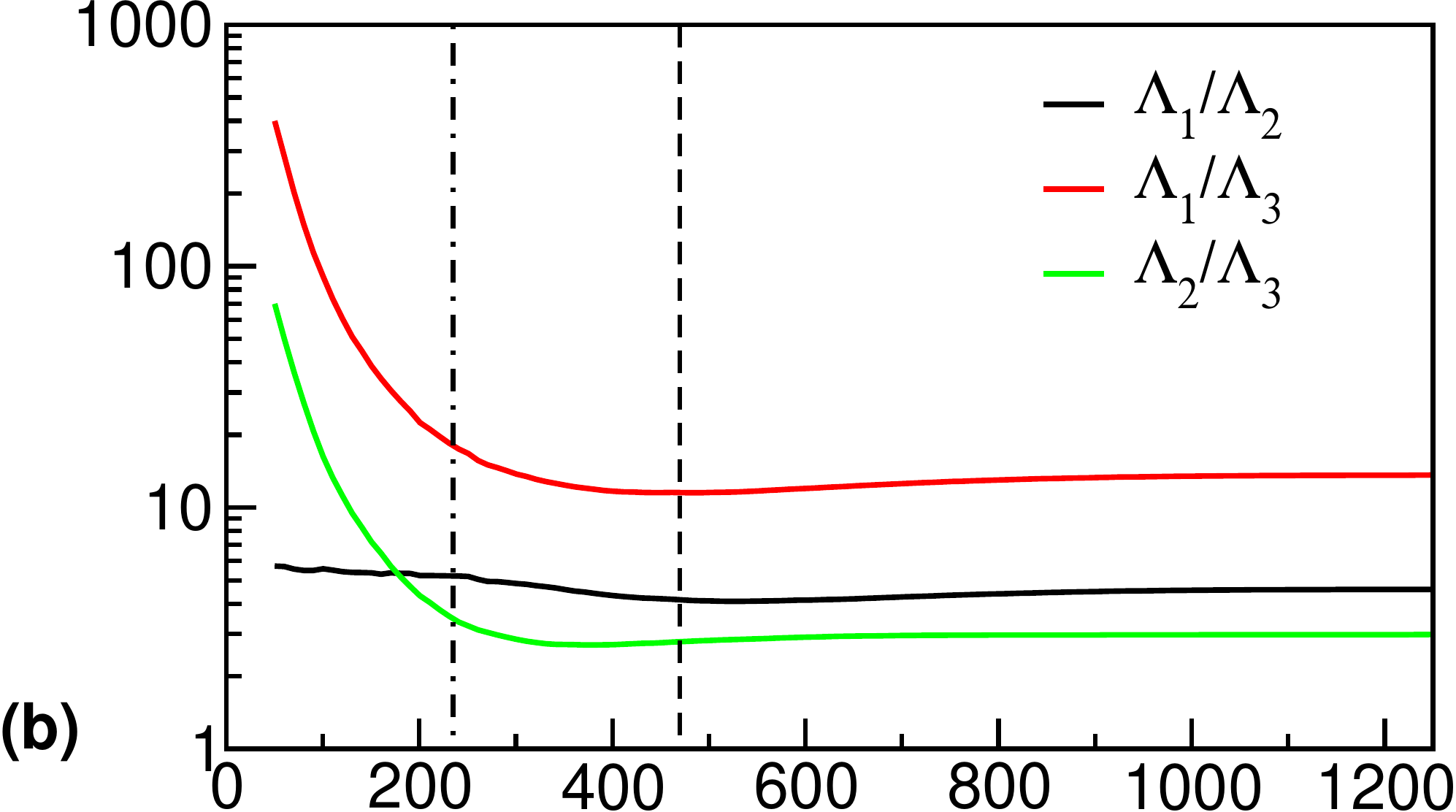}
}\\
\subfigure{
\includegraphics[width=3.0in]{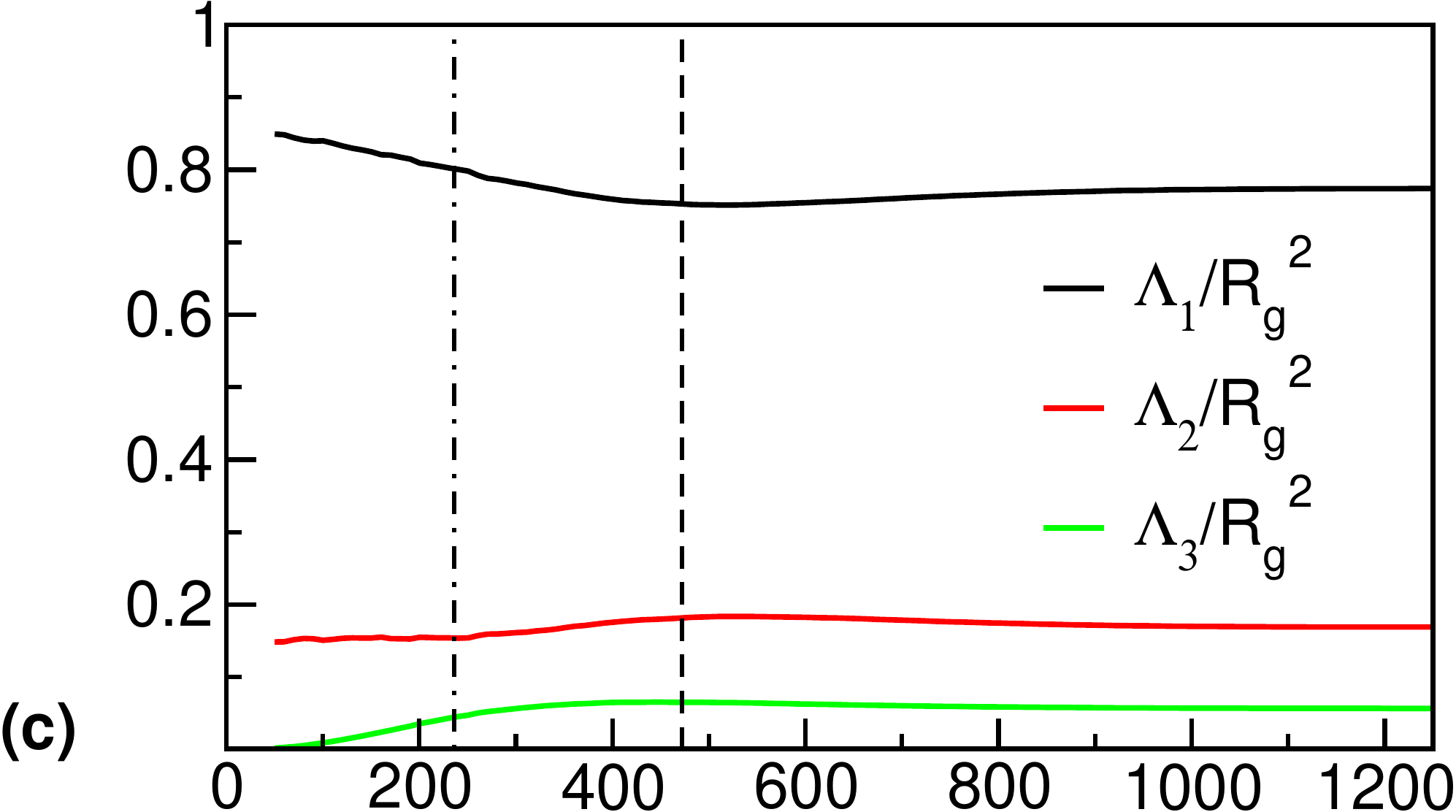}
}\\
\subfigure{
\includegraphics[width=3.0in]{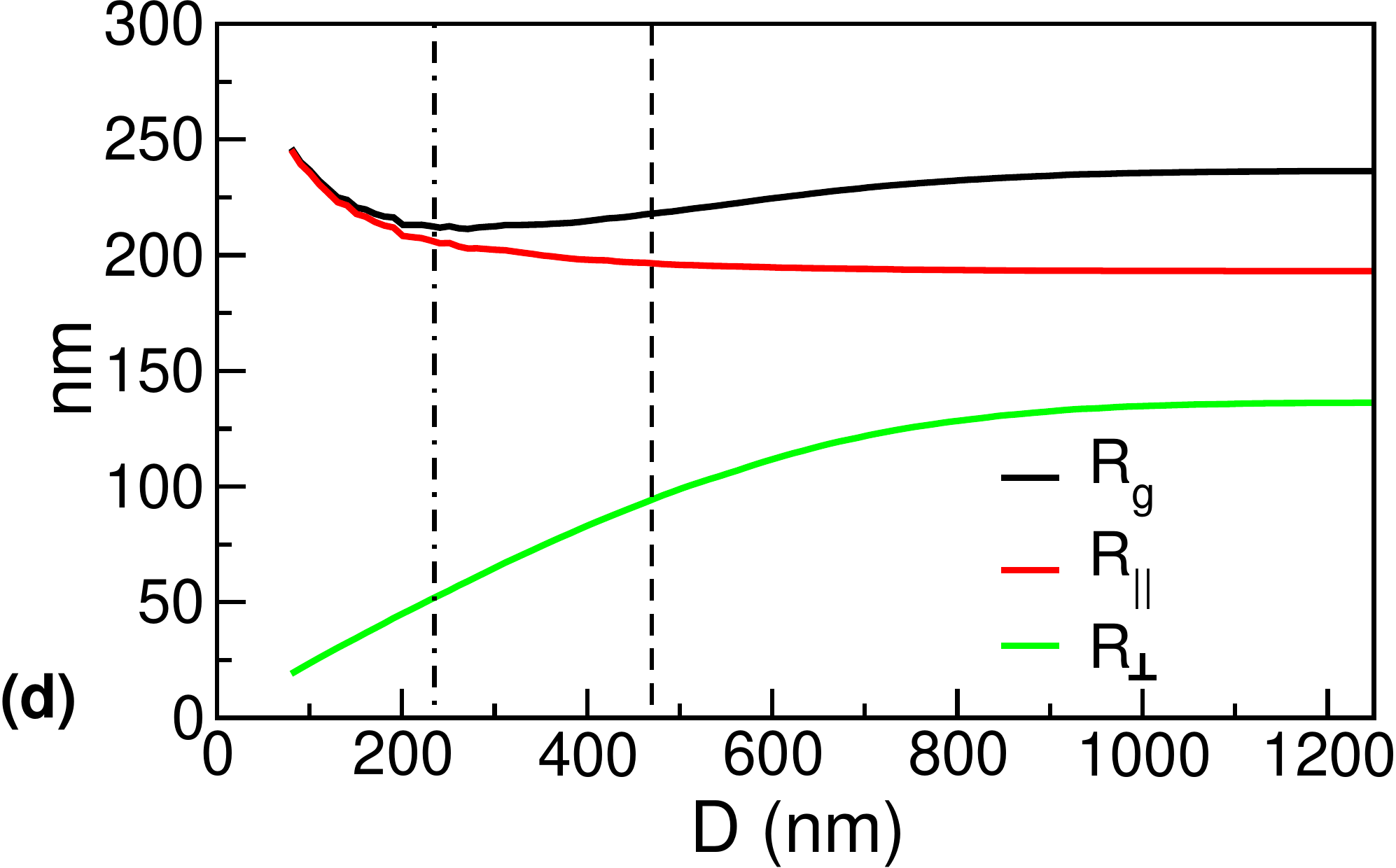}
}
\caption{Metric properties of a linear chain $L_c = 3604$ nm inside a
slit of width $D$. The orientation, shape and size of the gyration
ellipsoid of the confined chain are shown in panels (a--c). The root
mean square values of the chain radius of gyration, and its in- and
out-of-plane projections are shown in panel (d). The range and scale
of the $x$ axis (representing the slit width, $D$) is common to all
panels. The two vertical lines mark the reference slit widths $R_g^0$ (dot-dashed)
and $2 R_g^0$ (dashed).}
\label{fig:open_chain_orientation_and_size}
\end{figure}

The effect of slit confinement on the chain orientation is illustrated
in Fig.~\ref{fig:open_chain_orientation_and_size}a.

% Please arrange following figures in one single column (they share the x labels and tick marks)

The data reflect the in-plane and out-of-plane orientations, see
eq.~\ref{eqn:alphai}, of the three axes of inertia of open chains of
contour length, $L_c=3604$ nm. The bulk root mean square radius of
gyration of these chains is $R_g^0= 236$ nm (see Table
\ref{tab:isolchains}) and its location in the plots of
Fig.~\ref{fig:open_chain_orientation_and_size} is depicted by the
vertical dot-dashed line.  For slit widths much larger than the chain
bulk size the three mean square cosines are all close to the value
$1/3$, as expected in case of isotropic orientation of the chain in
space.

The orientation of the longest axis of inertia begins to be noticeably
non-isotropic when the slit width is about 700nm, which exceeds three
times the bulk radius of gyration. This is consistent with the strong
shape anisotropy of unconstrained chains ($\Lambda_1/\Lambda_3 = 13.8$
and $\Lambda_2/\Lambda_3 = 3.05$) which implies that the longest axis
of gyration cannot be freely-accommodated inside the slit already when
the width $D$ is several times larger than $R_g^0$. The second longest
axis of gyration is oriented isotropically, and
maintains its bulk value, down to slit widths $D
\approx 2 R_g^0 =$ 472nm (shown in the panels of
Fig.~\ref{fig:open_chain_orientation_and_size} as a dashed vertical
line).  The value of $\Lambda_2$ reduces appreciably when $D$ is
decreased down to about $R_g^0$ and remains approximately constant for
smaller slit widths. Analogous properties are found for the other
chain lengths considered in this study.

For $D < R_g^0$, both the first and the second axes of inertia are
mostly confined in the slit plane, while the third is oriented
perpendicularly to it. 

The progressive in-plane orientation of the two principal axes of
inertia is paralleled by the flattening of the chain in the slit
plane, see Fig.~\ref{fig:open_chain_orientation_and_size}b.  From
Fig.~\ref{fig:open_chain_orientation_and_size}c it is noticed that for
$D < 2R_g^0$ the progressive in-plane orientation of the second axis
of inertia is accompanied by a decrease of its eigenvalue $\Lambda_2$
to a limiting value that is smaller than in the unconstrained, bulk
case. The opposite is true for $\Lambda_1$. As a result, strongly
confined open chains tend to have an elongated shape, as in the
example of Fig.~\ref{fig:slab}a.

The adimensional data in
Fig.~\ref{fig:open_chain_orientation_and_size}b,c compares favourably
with available numerical and experimental results for the limiting
cases of no- and strong two-dimensional confinement.  In particular,
Haber et al.~\cite{Haber:2000:Proc-Natl-Acad-Sci-U-S-A:10984514} report the aspect ratios $\sqrt{\Lambda_1 / \Lambda_3} =
4.1$ and $\sqrt{\Lambda_1 / \Lambda_2} = 2.3$ for {\em fully-flexible}
unconstrained polymers, which are close to the corresponding values
3.7 and 2.1 found here. Finally, Maier {\em et al.}, in their study of
$\lambda$-phage DNA adsorbed on a flat bilayer lipid
membrane~\cite{Maier:2001:Macromol}, found $\Lambda_1/R^2_g = 0.84$ and
$\Lambda_2/R^2_g = 0.14$ which is in accord with the values $\Lambda_1/R^2_g = 0.84 \pm 0.02$ and $\Lambda_2/R^2_g = 0.148 \pm 0.008$ found here for the strongest two-dimensional confinement.

\subsection{Effect of confinement on chain size}

The global, in-plane and transverse size of the chain at various
levels of confinement are shown in
Fig.~\ref{fig:open_chain_orientation_and_size}d.

\begin{figure}[h!]
\centering
\subfigure{
\includegraphics[width=3.0in]{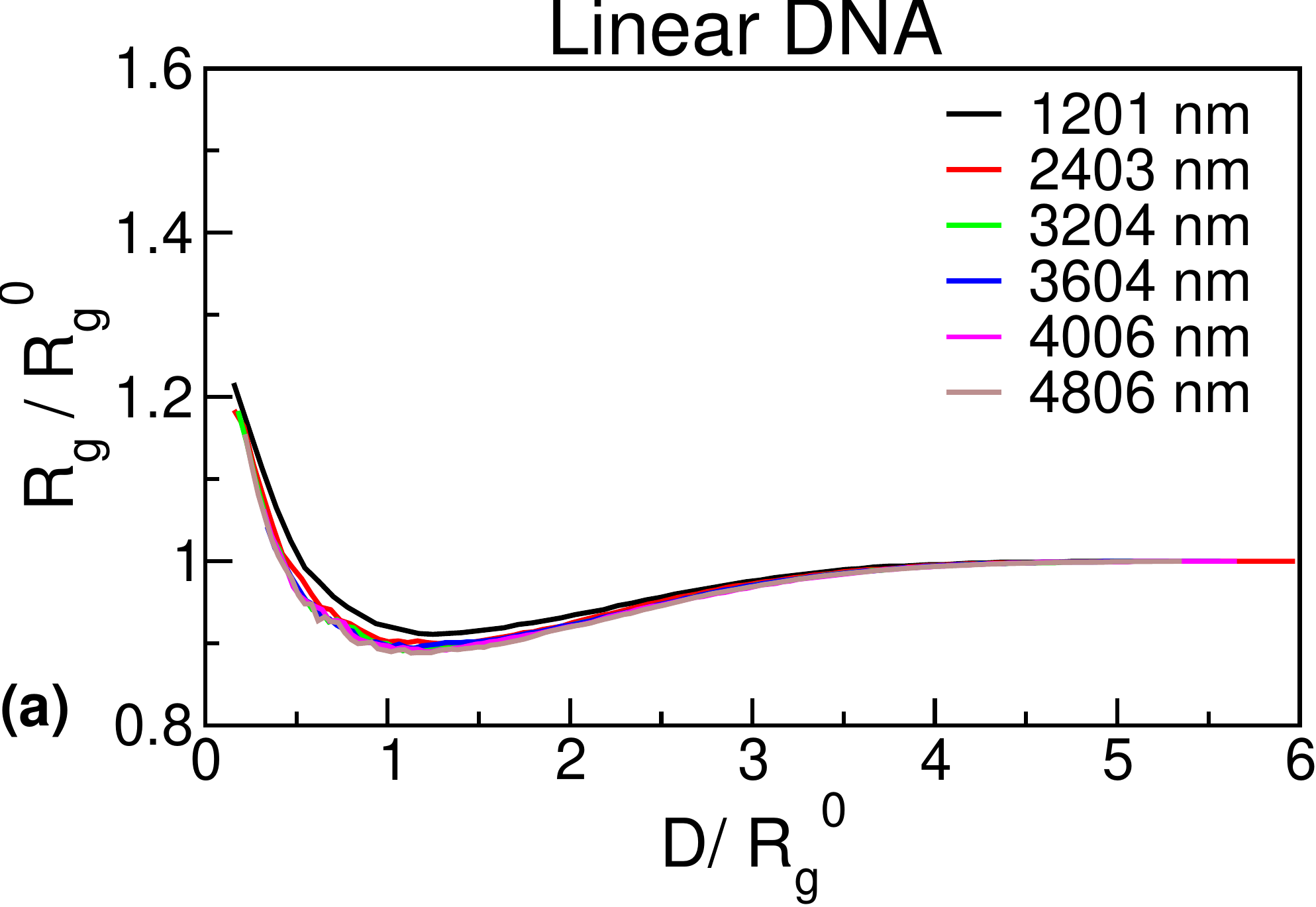}
}
\subfigure{
\includegraphics[width=3.0in]{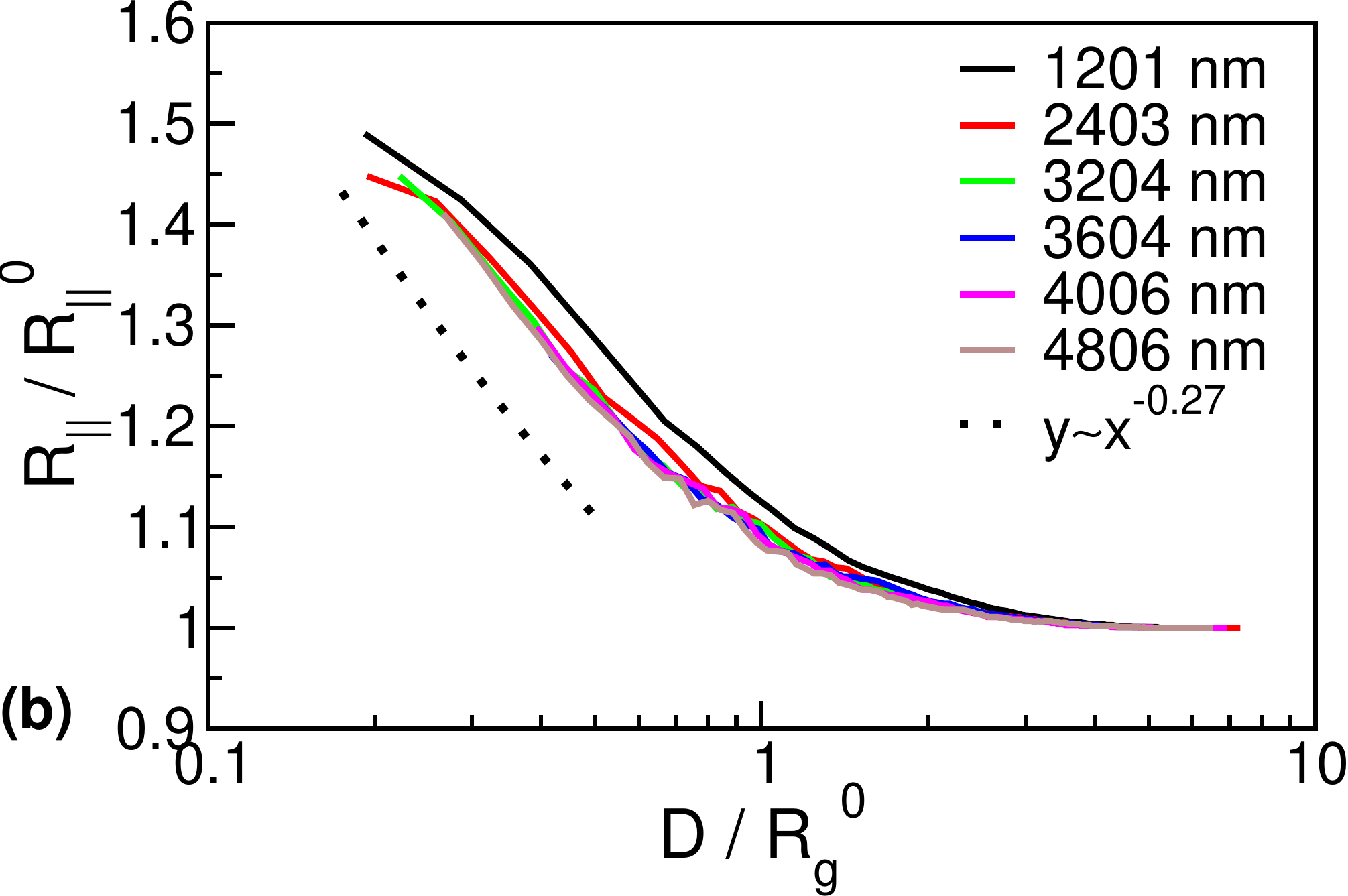}
}
\caption{(a) The minimum three-dimensional size of the chain is
  achieved when the slit width is about equal to the average radius of
  gyration of the unconstrained chain.  (b) Below such slit width the
  relationship between the average in-plane size, $R_{||}$, chain
  contour length, $N$, and slit width, $D$, is well accounted for by
  the DeGennes blob theory.}
\label{fig:chain_degennes}
\end{figure}

The transverse size, $R_{\perp}$ and the parallel one, $R_{||}$, have
opposite monotonic dependence on $D$ which result in a non-monotonic
behaviour of $R_g$ as a function of $D$.  From
Fig.~\ref{fig:open_chain_orientation_and_size}d, it is seen that $R_g$
is minimum for $D$ about equal to the bulk radius of gyration of the
chain, $R_g^0$. This property is found consistently across chains of
all considered lengths, as shown in Fig. \ref{fig:chain_degennes}a.

To test the validity of the de Gennes' blob regime we analyzed the
$D$ dependence of $R_{||}$. 
The result is shown in Fig.~\ref{fig:chain_degennes}b which reveals a remarkable collapse of
data points from curves of lengths $L_c$ between 2.4$\mu$m and
4.8$\mu$m.  As shown in the figure, for high confinement $R_{||}$ has
a power law dependence on $D$, with exponent equal to $-0.27 \pm
0.05$. This exponent, estimated by fitting the most asymptotic data
i.e. $L_c \ge 3604$ nm, is well compatible with the -0.276 value
predicted by the de Gennes' theory. Notice that, at all considered
lengths no deviation from the de Gennes' theory is observed,
consistently with the recent experiments of Doyle et
al.~\cite{Tang_et_al:2010:Macromol} on longer DNA filaments.

\section{Circular DNA in nano-slits}\label{sec:circular}

We consider DNA rings with the same contour length as the
previously-discussed open chains, see Table~\ref{tab:isolchains}, and
for the same range of confining slit widths. The properties of rings are first characterised at the metric level and then at
the topological one. We shall particularly focus on the type, abundance
and size of the various knots that are trapped in confined model DNA
molecules that are circularized in equilibrium. These aspects are
known to affect significantly the static and dynamics of polymers both in dilute\cite{Rawdon:Macromol:2008a,Rawdon:Macromol:2008b} and dense solutions~\cite{Rosa:2011:Macromol} or tightly packed in
small volumes~\cite{Arsuaga:2002:Proc-Natl-Acad-Sci-U-S-A:11959991,Matthews:2009:Phys-Rev-Lett:19257792,Marenduzzo:2009:Proc-Natl-Acad-Sci-U-S-A:20018693,Marenduzzo_et_al_2010_JoP_Cond_Matt,Min_entang_closure,Tubiana:2011:PRL}
or pulled through narrow
channels~\cite{Matthews:2009:Phys-Rev-Lett:19257792,Marenduzzo:2009:Proc-Natl-Acad-Sci-U-S-A:20018693}
and are hence expected to be important for two-dimensional confinement
too.

\subsection{Effect of confinement on ring orientation and size}

\begin{figure}[h!]
\centering
\subfigure{
\includegraphics[width=3.0in]{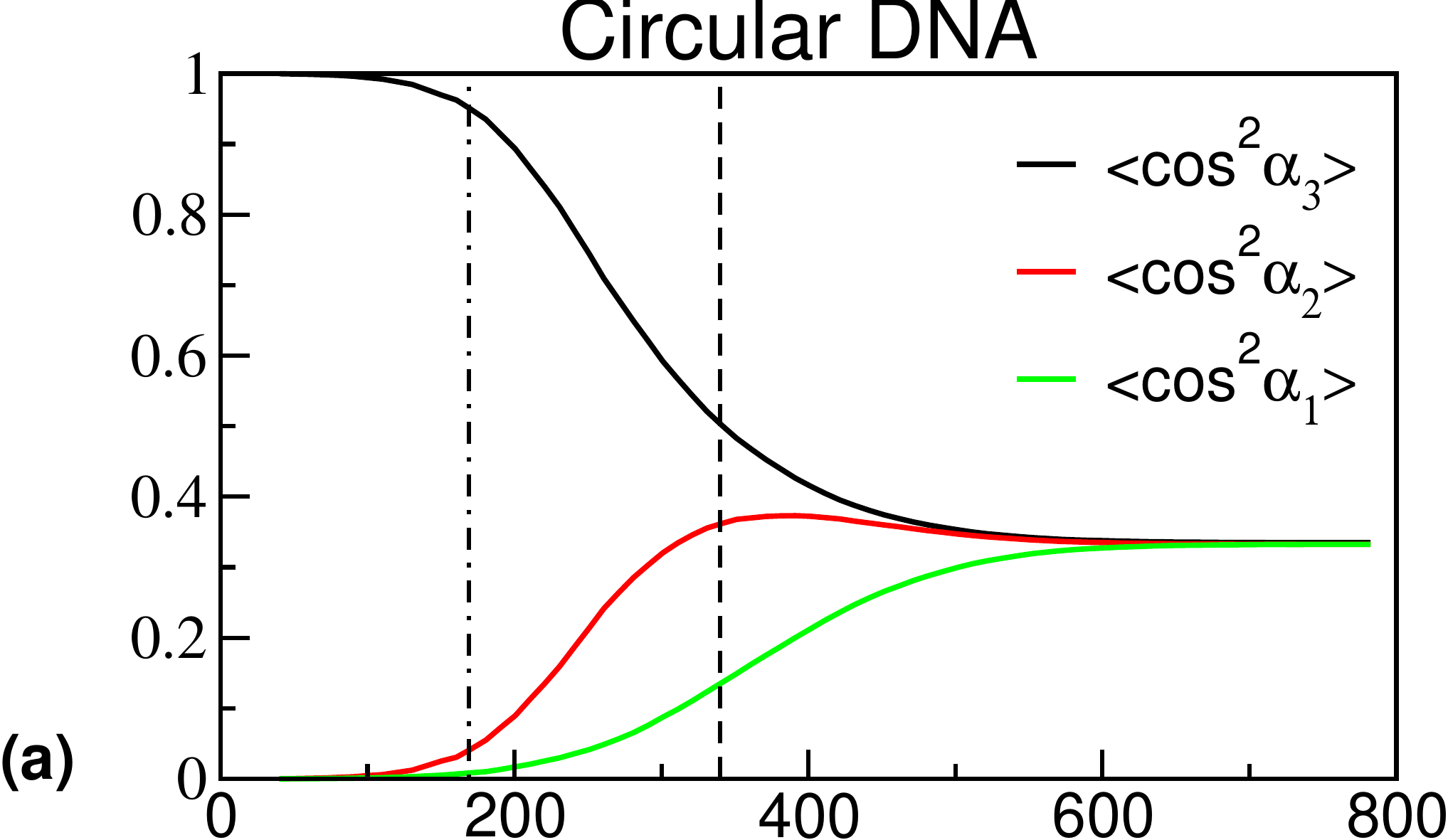}
}\\
\subfigure{
\includegraphics[width=3.0in]{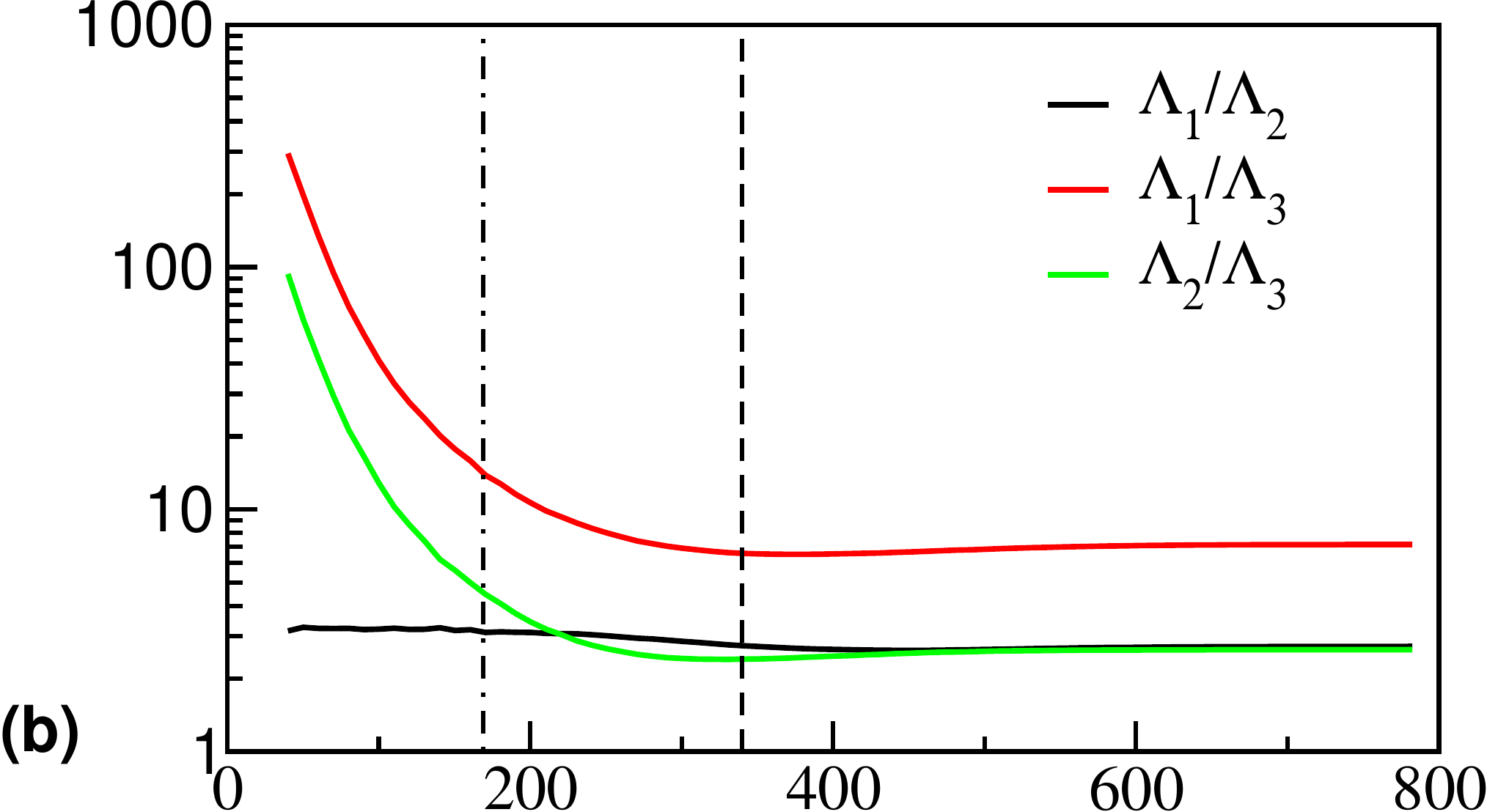}
}\\
\subfigure{
\includegraphics[width=3.0in]{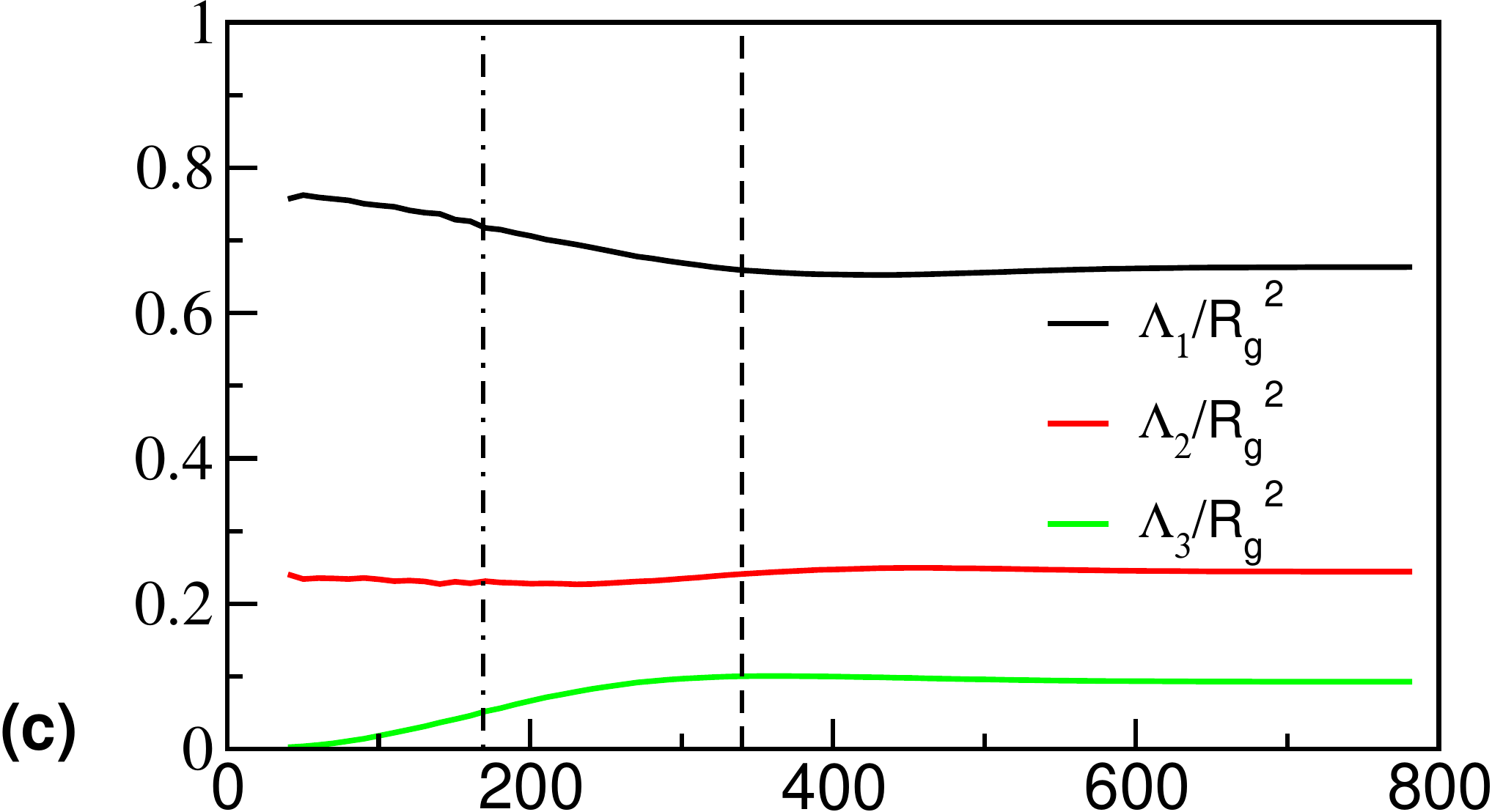}
}\\
\subfigure{
\includegraphics[width=3.0in]{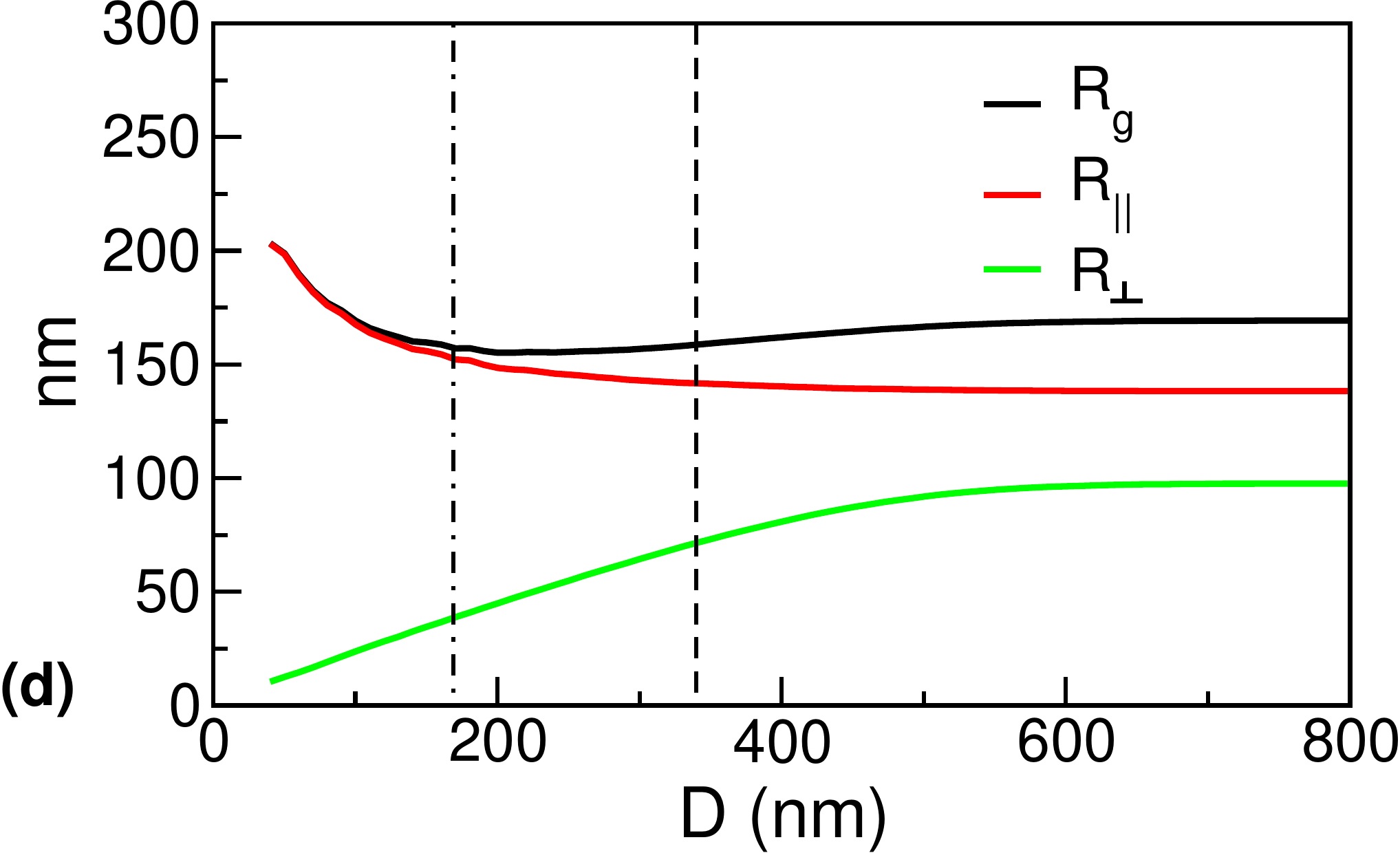}
}\\
\caption{Metric properties of a circular chain $L_c = 3604$ nm inside a
slit of width $D$. The orientation, shape and size of the gyration
ellipsoid of the confined ring are shown in panels (a--c). The root
mean square values of the ring radius of gyration, and its in- and
out-of-plane projections are shown in panel (d). The range and scale
of the $x$ axis (representing the slit width, $D$) is common to all
panels. The two vertical lines mark the reference slit widths $R_g^0$ (dot-dashed)
and $2 R_g^0$ (dashed).}
\label{fig:ring_orientation_and_size}
\end{figure}

The dependence of a ring size and orientation on the confining slit
width is illustrated in Fig.~\ref{fig:ring_orientation_and_size}.  The
data pertains to rings of length $L_c=3604$ nm and therefore can be
directly compared with those in
Fig.~\ref{fig:open_chain_orientation_and_size} for an equally-long
open chain.

% Please arrange following figures in one single column (they share the x labels and tick marks)

The comparison of panels (b) and (c) in the two figures indicates that
the shape of closed chains is significantly more spherical than open
ones at all levels of confinement. The effect is aptly illustrated by
the typical open and closed chains shown in Fig.~\ref{fig:slab}.

Notwithstanding the different shape, it is found that the changes in
the ring orientation and size occur for values of the normalized slit
width, $D/R_g^0$ that are similar to the linear case. In particular,
the data in panel (a) show that the in-plane orientation of the ring
principal axis of gyration sets in for $D\approx 2 R_g^0$ while for
the second one occurs at $D\approx R_g^0$. Furthermore the average
ring size, $R_g$, shown in panel (d) has a non-monotonic dependence on
$D$ and its minimum value (again given by the opposite trends of
$R_{||}$ and $R_{\perp}$) occurs for $D \approx R_g^0$, as for the
linear case.

\subsection{Scaling properties and the deGennes regime}

Because the bulk metric scaling exponents $\nu_{\rm 2D}$ and $\nu_{\rm 3D}$
are the same for open and closed chains, it is expected
that the de Gennes scaling regime, originally based on the blob
argument for linear chains, ought to hold for circular ones too.

However, the closed character of the rings introduces correlations in
the orientation in pairs of bonds at a large chemical (contour)
distance that have no analog in linear chains~\cite{Witz:2008:PRL,PhysRevLett.106.248301}

\begin{figure}[h!]
\centering
\subfigure{
\includegraphics[width=3.0in]{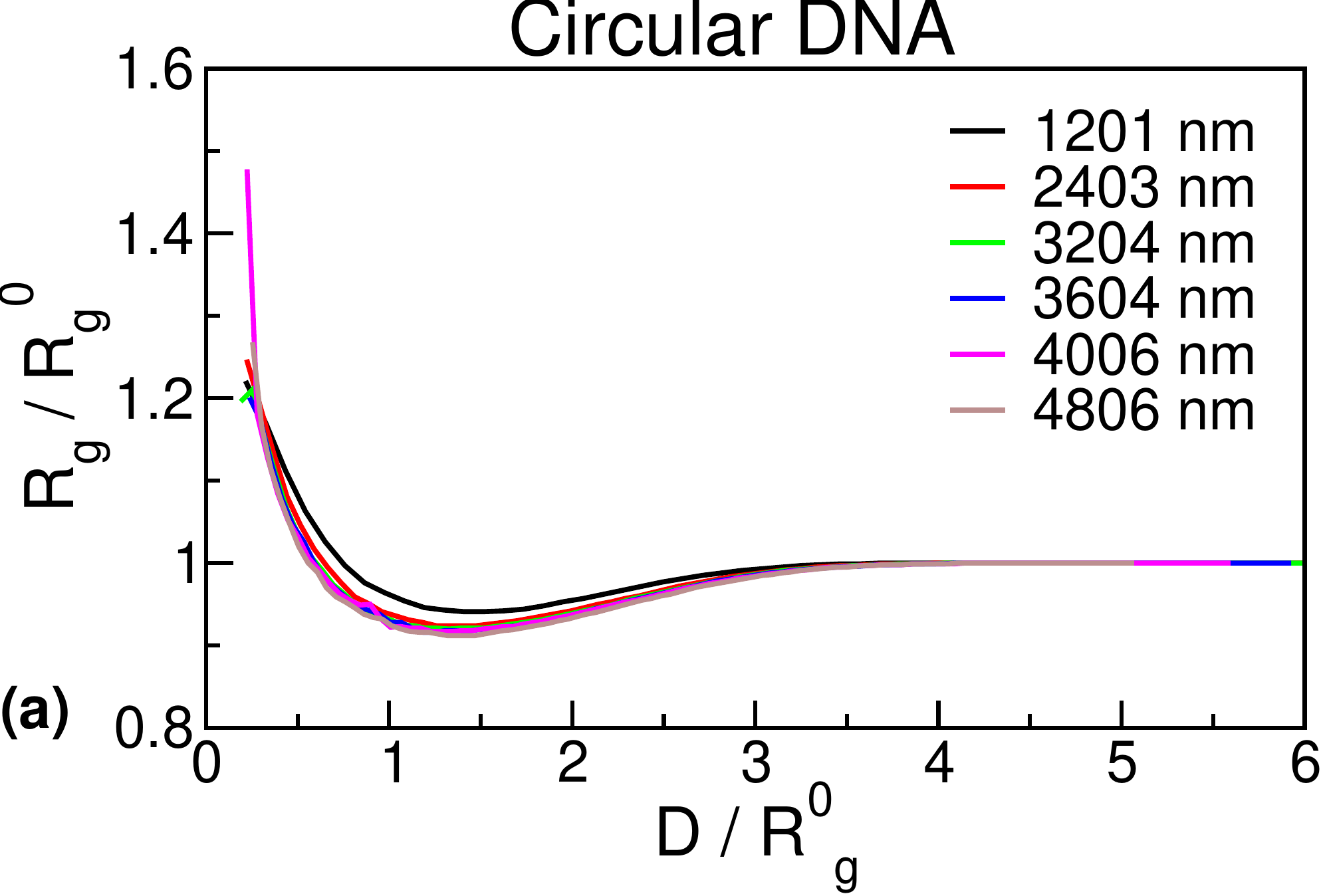}
}
\subfigure{
\includegraphics[width=3.0in]{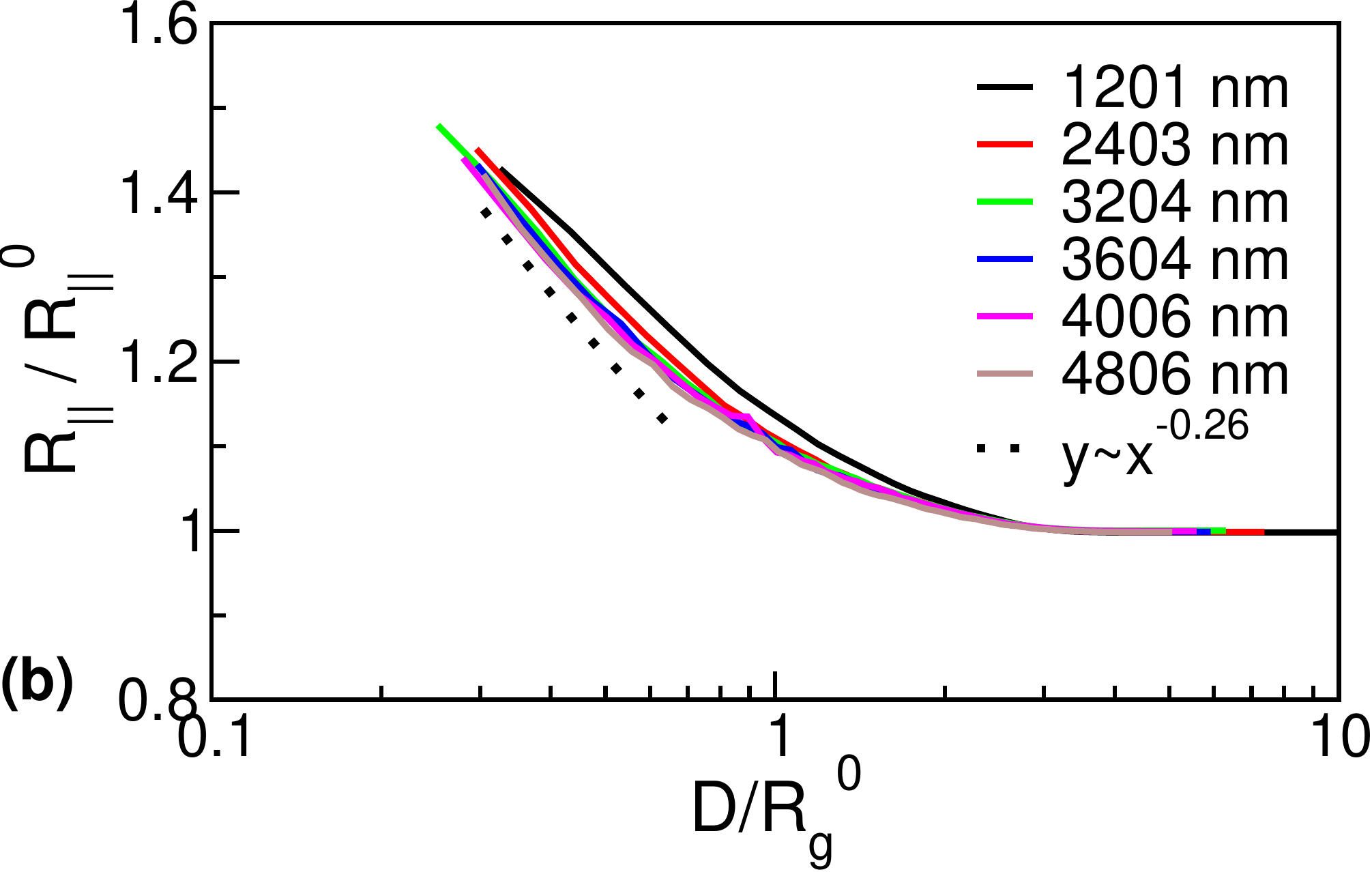}
}
\caption{(a) The minimum three-dimensional size of the ring is
achieved when the slit width is about equal to the average radius of
gyration of the unconstrained ring.  (b) Below such slit width the
relationship between the average in-plane size, $R_{||}$, ring contour
length, $N$, and slit width, $D$, is well accounted for by the de
Gennes' blob theory.}
\label{fig:ring_degennes}
\end{figure}

To clarify the effect we show in Fig.~\ref{fig:ring_degennes} the
same observables reported in Fig.~\ref{fig:chain_degennes} to analyze
the de Gennes regime for confined linear chains.

Panel (a) shows that, except for the data at the smallest ring contour
length, there is a good overlap of the rescaled ring size versus width
curves. In particular, for all considered chain lengths the minimum
overall ring size is obtained for $D \approx R_g^0$, consistently with
the open chain case

For $D < R_g^0$ the mean in-plane size shows a behaviour that is
compatible with the de Gennes scaling regime, as illustrated in panel
(b). The power-law fit of the asymptotic data in the figure, i.e. $L_c
\ge 3604$nm and $D/R_g^0 < 1$, yields the scaling exponent $-0.26 \pm
0.04$. This value is compatible with the -0.276 value predicted by
the de Gennes theory, consistently with what found for the linear case.

\subsection{Topological entanglement}

\subsubsection{Knotting probability}

The Monte Carlo strategy employed here and in other previous studies provides
the means to characterize the occurrence of non-trivial knots in rings
that are circularised in equilibrium. The knotting probability of
unconstrained rings depends on several lengthscales such as the
molecule's persistence length, its contour length and
thickness~\cite{Frank-Kamenetskii:1975:Nature,Deguchi&Tsurusaki:1997:PRE}. 
Accordingly, the topological complexity of the circularised
molecules can be tuned, to some extent, by varying these lengths, for
example by intervening of the solution ionic
strength~\cite{Rybenkov:1993:Proc-Natl-Acad-Sci-U-S-A:8506378,Toan:2006:JPhysCondMat}.

Spatial confinement introduces a further lengthscale in the problem,
the width of the confining region, which can affect dramatically the
molecule's entanglement. The effect has been addressed by several
studies of isotropic three-dimensional confinement~\cite{Arsuaga:2002:Proc-Natl-Acad-Sci-U-S-A:11959991,Micheletti:2006:J-Chem-Phys:16483240,Micheletti:2008:Biophys-J:18621819,Marenduzzo_et_al_2010_JoP_Cond_Matt}  which
established that progressive reduction of the volume resulted in an
ever-increasing fraction of knotted molecules and of their geometrical
and topological complexity~\cite{Tubiana:2011:PRL}.

By comparison, the case of two-dimensional confinement is virtually
unexplored, except for lattice polymer
models~\cite{Whittington&Soteros:1991:IJC,Tesi:1994:J-Phys-A,Janse-van-Rensburg:2007:JSTAT}.
Making progress in this problem is not only important to fill a
conceptual gap but is especially so in view of the dramatic
qualitative differences that are expected {\em a priori} compared
to the isotropic, three-dimensional, confinement.

In fact, it should be considered that for extreme two dimensional
confinement, i.e. for slit width only slightly larger than the chain
intrinsic thickness, over/under-crossings of the chain are impossible
and knots cannot be formed. Therefore, unlike the isotropic case,
progressive slit confinement cannot result in an ever increasing
complexity of the ring topology and geometry. 

\begin{figure}[h!]
\includegraphics[width=3.0in]{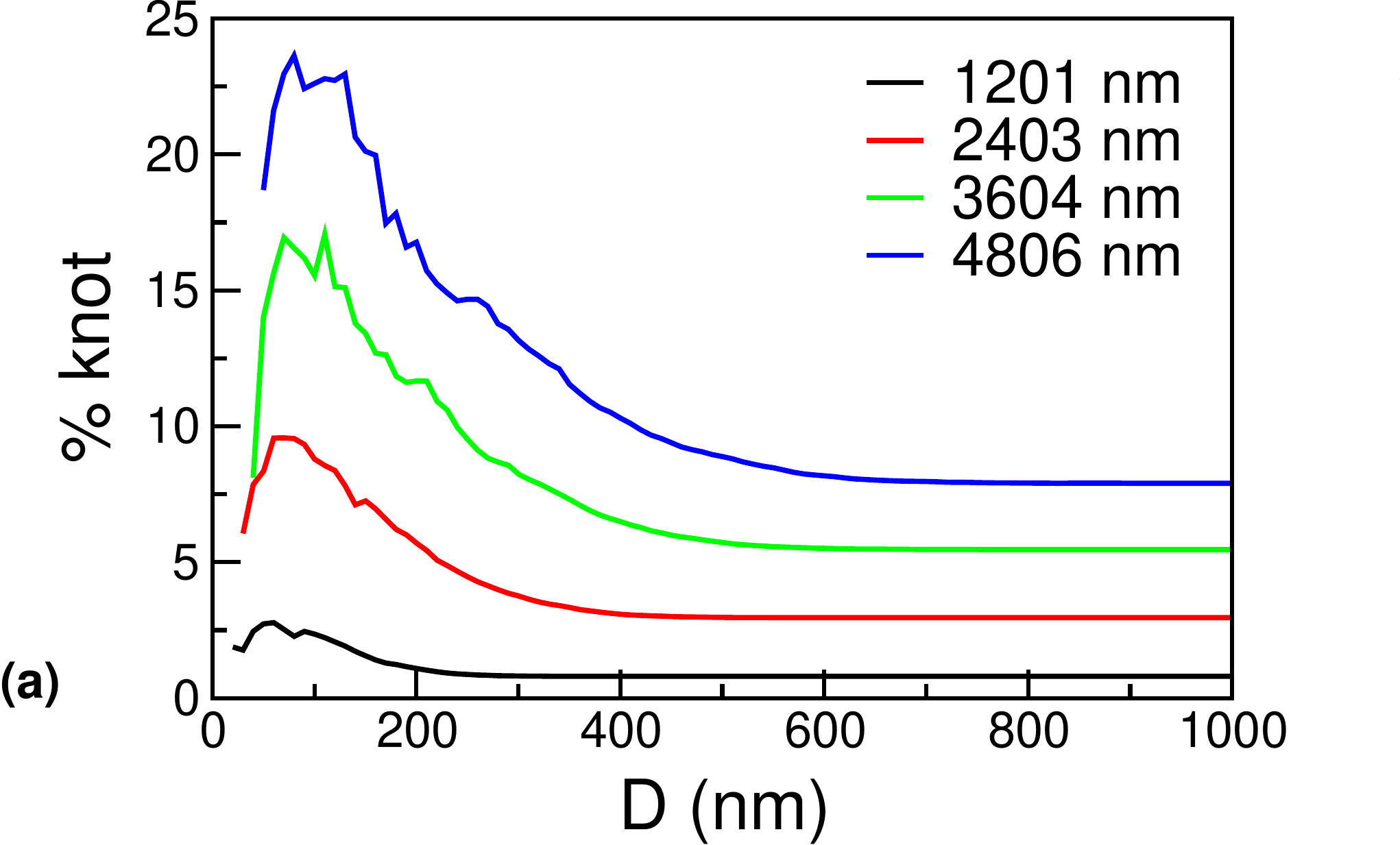}
\includegraphics[width=3.0in]{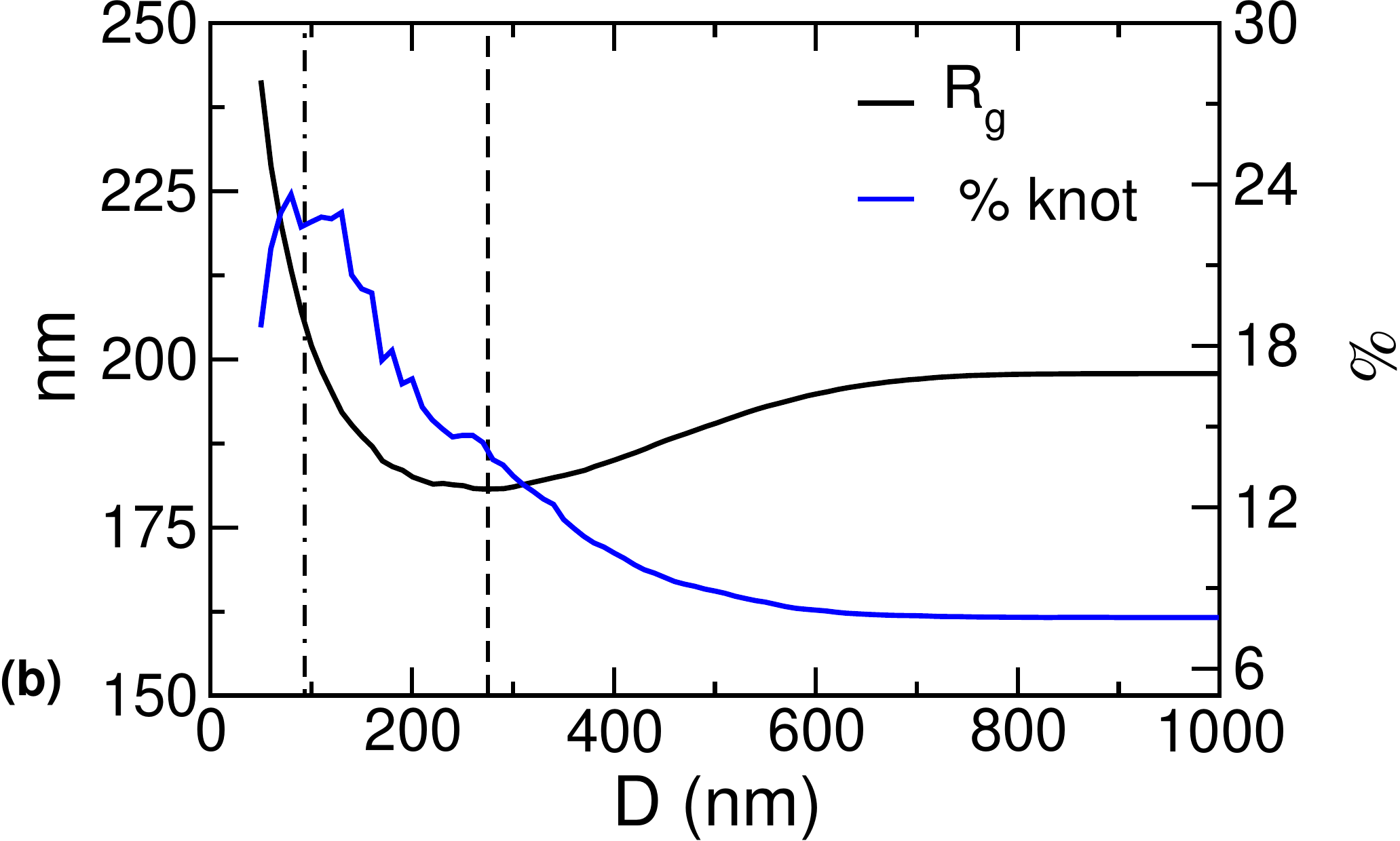}
\caption{(a) Knotting probability of equilibrated rings of different
  contour length. (b) Dependence on the slit width, $D$, of the
  average radius of gyration and knotting probability for a circular
  DNA ring of $L_c=4806$nm.  The vertical dashed line marks the slit
  width, $D^*$, corresponding to the minimum three-dimensional size of
  the ring.  The dot-dashed line marks the maximum of the knotting
  probability.}
\label{fig:knotting}
\end{figure}

To clarify this aspect we investigated the dependence of the knotting
probability on $D$. The results are shown in Fig.~\ref{fig:knotting}a
and indicate that the ring knotting probability is non-monotonic and
for a certain slit width, $D_e$ reaches a maximum enhancement compared
to the bulk case. Notice that the peak value exceeds the bulk knotting
probability by several factors. The figure indicates that the limiting
value of zero knotting probability is rapidly reached for $D < D_e$.

It could be intuitively expected that the maximum knotting enhancement
occurs for the same slit width, $D^*$, associated to the maximum chain
density (minimum $R_g$)~\cite{Tesi:1994:J-Phys-A}. This is, however,
not the case. In fact, in the example of Fig.~\ref{fig:knotting}b the
peak of the probability distribution occurs when the slit width is
$D_e\approx 90$nm, which is about one third of $D^*$.  As a matter of fact, over the
considered range of contour lengths, $D_e$ is consistently equal to
about half the bulk ring size.

\subsubsection{Knot spectrum}

It is most interesting to examine how the various, topologically
inequivalent knot types, contribute to the nonmonotonic trend of the
overall knotting probability.

\begin{figure}[h!]
\includegraphics[width=3.in]{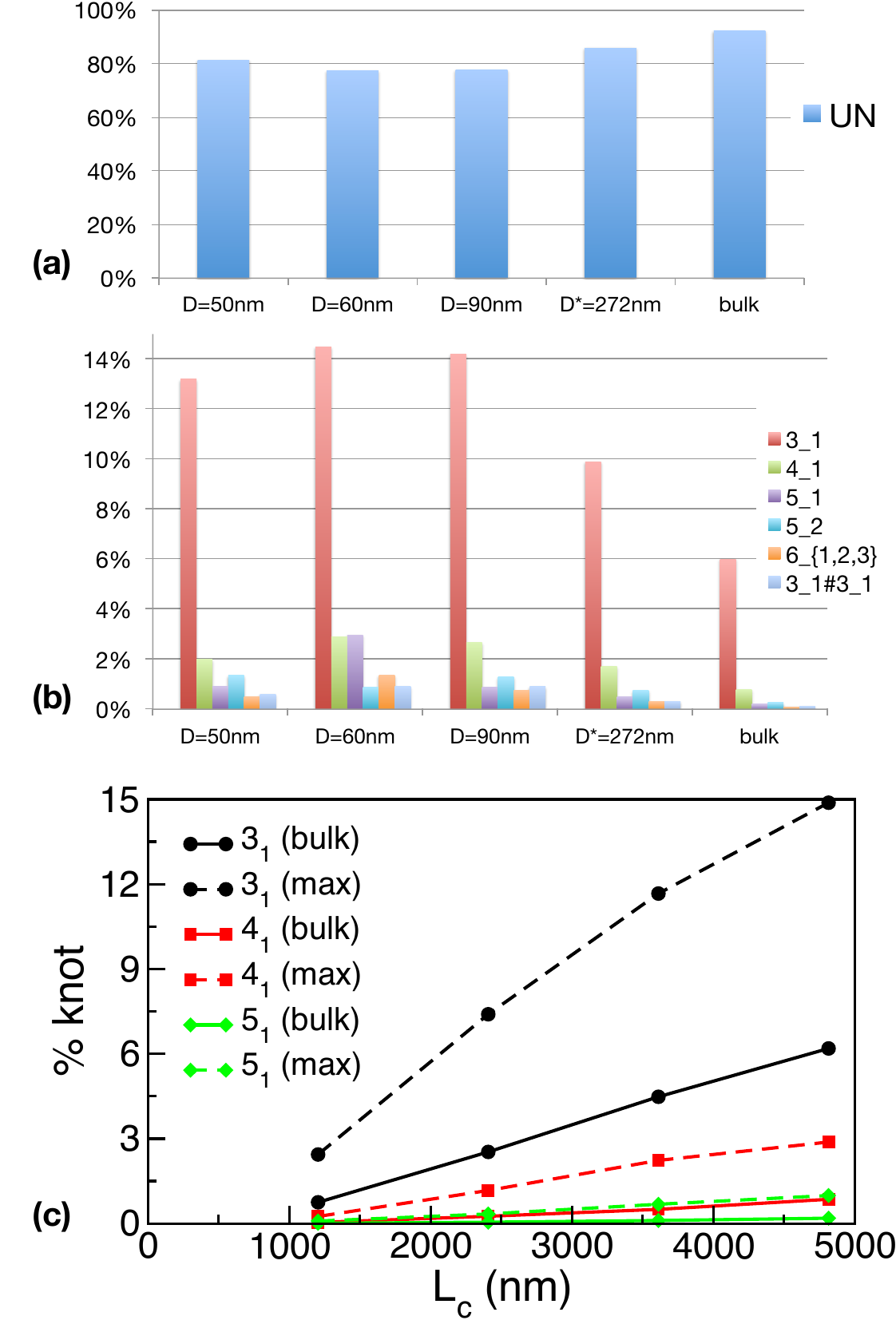}
\caption{(a,b) Slit-width dependence of the knot spectrum of confined
circular DNA with $L_c=4806$nm. $D^*$ is the slit width at which the
mean 3D extension of the rings reaches its minimum value.  (c) Maximun
probability of occurrence for knot types $3_1$, $4_1$ and $5_1$ as a
function of the contour length $L_c$ (dashed curves). Solid curves
refer to the corresponding case of unconstrained circular DNAs.}
\label{fig:spectrum_and_enhancement}
\end{figure}

The results are shown in Fig.~\ref{fig:spectrum_and_enhancement}a,b
which portrays the probability of occurrence of various knot types
(including the unknot) for four different slit widths and
$L_c=4.8\mu$m. It is seen that, at all level of confinement, the knot
population consists of very simple knots, and is dominated by the
simplest of all, the trefoil or $3_1$ knot, whose peak abundance is
$\sim$14\%. 

The result is notable in several respects. First, the knot spectrum of
Fig.~\ref{fig:spectrum_and_enhancement}a is qualitatively very
different from the one of an equivalent model of circular DNA subject
to isotropic, spherical confinement~\cite{Micheletti:2008:Biophys-J:18621819}.  

In fact, even for moderate three-dimensional confinement in a sphere
of diameter equal to 180nm, the knot spectrum of molecules of
$Lc=3.4\mu$m is largely constituted by highly complex knots: trefoils
are less than 10 \% of the non-trivial knots (which, in turn are 55\%
of the ring population). Conversely, for slit confinement, at all
considered values of $L_c$ and $D$, complex knots are virtually absent
and the population of simple knot types, such as the trefoil, reaches
values that cannot be matched by rings of the same or longer contour lengths subject to any degree of isotropic
confinement~\cite{Micheletti:2006:J-Chem-Phys:16483240}.

We finally turn to the dependence of the peak of the 
probability of
occurrence of simple knots as a function of chain length. The data in
Fig.~\ref{fig:spectrum_and_enhancement}c indicate a monotonic increase
of the peak probability with $L_c$. Therefore, the population of
simple knots can be further increased compared to the values in panel
(b) by suitably confining rings with $L_c > 5\mu$m. At the same time,
it is noticed that the derivative of the peak probability curves
decreases with $L_c$ so that, by analogy with the bulk probability
case, one can envisage that the curves will eventually decrease for
sufficiently large values of $L_c$. This is an interesting point that
would deserve to be addressed in future studies.

\subsubsection{Knot length}

\begin{figure*}[h!]
\includegraphics[width=6.0in]{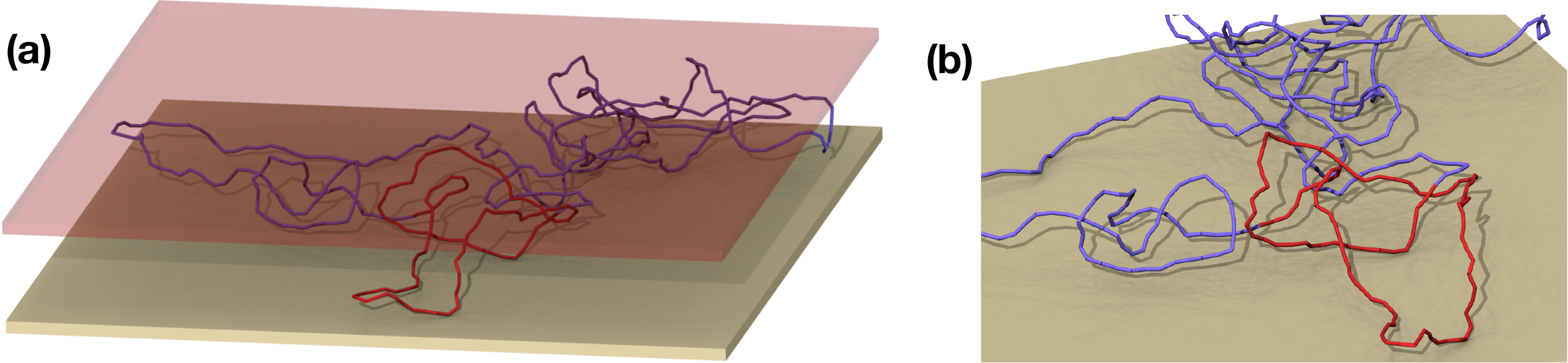}
\caption{(a) Trefoil-knotted ring confined in a nanoslit. The
  configuration is the same as the one shown in Fig. \ref{fig:slab}c,d. The ring
  contour length is 0.48 $\mu$m and the slit width is 70nm. The
  shortest knotted portion of the ring is highlighted in red. (b)
  Close-up of the knotted portion; for graphical clarity the slit top is
  not shown.}
\label{fig:fig_3_1_P08}
\end{figure*}

We conclude the topological characterization by reporting on the
degree of localization (in space and along the chain) of the knots
formed in the confined rings. The knot location along the ring was
established using the bottom-up search of the smallest arc(s) that,
after closure, has the same topology of the whole
ring~\cite{Marcone:2005:J-Phys-A,Min_entang_closure}, see the example
in Fig.~\ref{fig:fig_3_1_P08}.

\begin{figure}[h!]  
\includegraphics[width=3.0in]{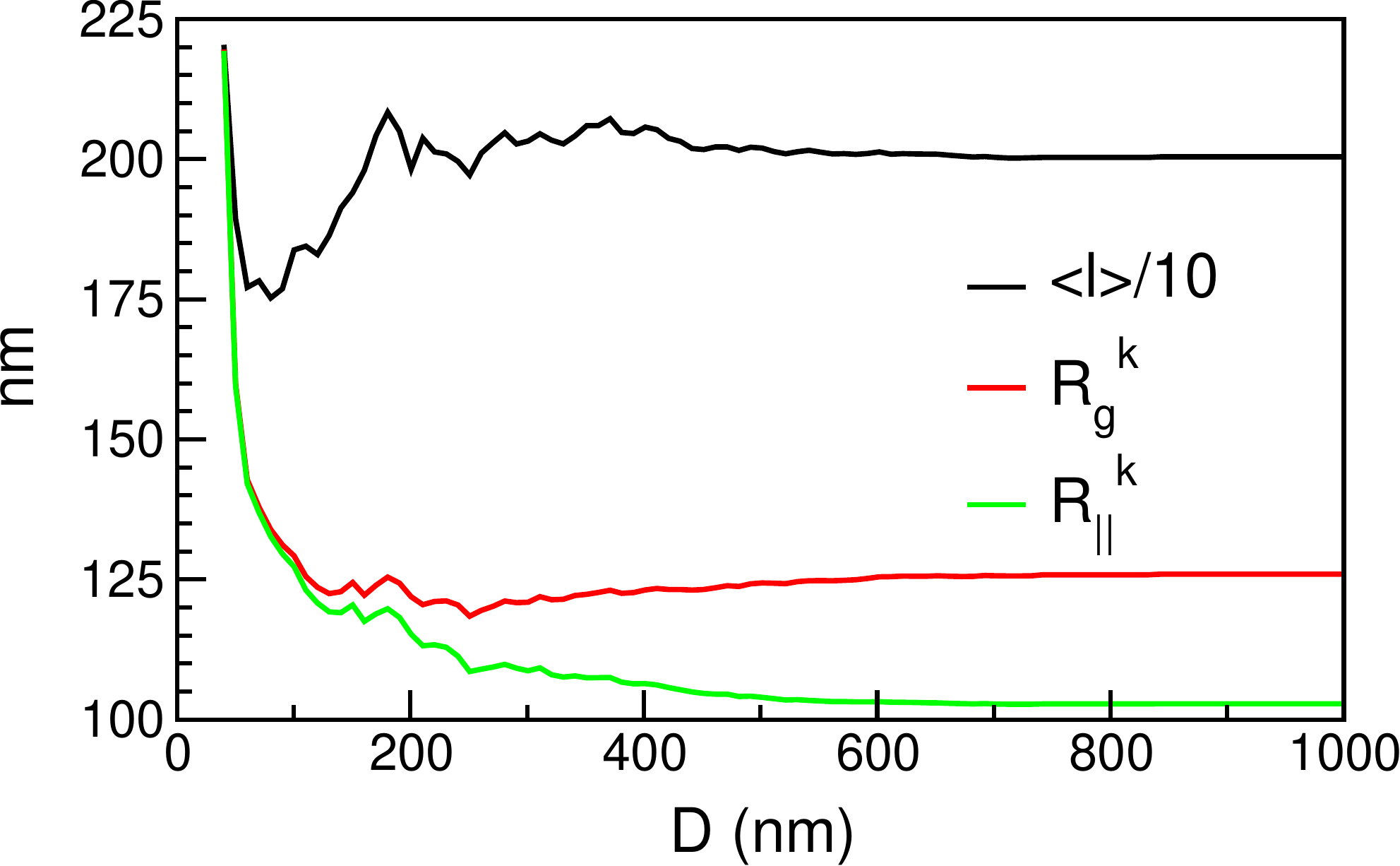}
\caption{Three-dimensional and parallel mean estension of the knotted
  portion of circular trefoil knots with $L_c = 4806$nm.  Circles refer
  to the average knot size expressed in unit of cylinders.  }
\label{fig:3_1_size_metric_N480}
\end{figure}

The average contour length spanned by the knots (all non-trivial
topologies) as a function of $D$ is shown in
Fig.~\ref{fig:3_1_size_metric_N480} for rings of length
$Lc=4.8\mu$m. It is seen that up to the point where the average ring
size is minimum, the length of the knotted region is about constant,
and drops significantly at smaller slit widths.

The same figure also shows the average in-plane and overall size of
the knotted region, which is indicated with $R_g^k$ and $R_{||}^k$,
respectively. It is apparent that the drop in knot length occurs when
the slit width is about equal to the bulk value of the overall knot
size, $D\approx 125$nm. These results suggest that
for knotted polymer rings confined into slits the typical size of the
knotted region introduces an additional length scale
  into the problem that could play a relevant role in understanding
  the metric and mechanical properties of these confined molecules at
  equilibrium~\cite{Janse-van-Rensburg:2007:JSTAT,Matthews_et_al:2011:Mol_Phys}.

\section{Conclusions}\label{sec:conclusions}

The equilibrium properties of linear and circularised model DNA molecules
confined in nano-slits were studied by numerical methods. By using
thermodynamic reweighting techniques the properties of these systems
were systematically profiled as a function of the molecules' contour length and
nano-slit width. The analysis proceeded by first considering the
molecules' geometrical properties and next the topological ones. The
former included the size, shape and spatial orientation of the
molecules in the slit, while the latter consisted of the knotting
probability and knot spectrum.

We found that the scaling of the metric properties was described very
well by de Gennes' blob theory for both linear and circular molecules
of all considered lengths and down to the highest confinement. To the
best of our knowledge, it is the first time that the applicability of
the blob theory is discussed (and shown to hold) for equilibrated ring
polymers. In addition, even for linear molecules, the fact that the de
Gennes' scaling is investigated upon varying {\em both} the molecules'
contour length and the slit width represents a significant broadening
of the scope of previous investigations. Our results are consistent
with the conclusions of recent experiments of nano-slit confinement of
linear $\lambda$-DNA  which found no indications of deviations
from de Gennes' blob theory~\cite{Tang_et_al:2010:Macromol}.

Finally, and most importantly, we characterized the topological
entanglement that is trapped in  molecules that
circularise in equilibrium inside nano-slits. For both applicative purposes and for
fundamental polymer science, the characterization of the physical
properties of DNA molecules with various topologies (linear, circular
planar, circular knotted) has attracted much interest in the past
decades. Our contribution represents a step forward in the
characterization of such systems in the presence of confinement: while
the three-dimensional confinement has been heavily
investigated~\cite{Arsuaga:2002:Proc-Natl-Acad-Sci-U-S-A:11959991,Arsuaga:2005:Proc-Natl-Acad-Sci-U-S-A:15958528,Micheletti:2006:J-Chem-Phys:16483240,Micheletti:2008:Biophys-J:18621819,Marenduzzo:2009:Proc-Natl-Acad-Sci-U-S-A:20018693},
the two-dimensional one by means of slits is still largely unexplored.
Our study has pointed out a remarkable feature that has no parallel
with three-dimensional confinement, namely the non-monotonicity of the
knotting probability for increasing confinement. Specifically, it is
seen that, among the rings resulting from equilibrium
circularization (studied without enforcing topology conservation) the
fraction of knotted rings has a maximum at a length-dependent slit
width, $D_e$. The peak knotting probability exceeds by several
factors the one in the bulk. Upon decreasing $D$ below $D_e$, the
knotting probability decreases very rapidly to zero.

The above result shows that nano-slit confinement ought to be
profitably used in contexts where an {\em a priori} control of the
knotted topology of DNA rings is desired. In particular, the
non-monotonic dependence of the knotting probability as a function of
confinement, makes it possible to conceive experiments where the slit
width is tuned so that the knotting probability is either
significantly enhanced or largely depressed with respect to the bulk case.

The natural extensions of the present study are twofold. On the one
hand it would be interesting to consider much longer chain lengths and
establish the detailed length-dependence of $D_e$ and the associated
maximum knotting probablity.  On the other hand, it would be most
interesting to repeat the present investigation to the one-dimensional
confinement case (i.e. nano-channels) and analyze how the expected
consecutive onsets of the de Gennes and Odijk regimes impact both the
geometrical and the topological properties of confined rings.

\section{Acknowledgements}
We acknowledge support from the Italian Ministry of Education. We thank A. Rosa for useful discussions.
\vskip 0.5cm
Reprinted with permission from\\
C. Micheletti and E. Orlandini, Macromolecules, 2012, vol. 45, pp. 8668-8680.\\
Copyright (2012) American Chemical Society.

\end{document}